\documentclass[pdflatex,sn-mathphys-num]{sn-jnl}

\usepackage{graphicx}%
\usepackage{multirow}%
\usepackage{amsmath,amssymb,amsfonts}%
\usepackage{amsthm}%
\usepackage{mathrsfs}%
\usepackage[title]{appendix}%
\usepackage{xcolor}%
\usepackage{textcomp}%
\usepackage{manyfoot}%
\usepackage{booktabs}%
\usepackage{listings}%
\usepackage[ruled,vlined,linesnumbered]{algorithm2e}
\usepackage{algorithmic}
\usepackage{amsfonts}
\usepackage{amssymb}
\usepackage{caption}
\usepackage{tablefootnote}
\usepackage{longtable}
\usepackage{hhline,multirow}
\usepackage{rotating}
\usepackage{mathtools}
\usepackage{url}

\usepackage{breakurl}
\theoremstyle{thmstyletwo}%

\theoremstyle{thmstylethree}%

\begin{document}

\title[Article Title]{Guarding Digital Privacy: Exploring User
Profiling and Security Enhancements}

\author*[1]{\fnm{Rishika} \sur{Kohli}}\email{rishika.kohli@iitjammu.ac.in}

\author[2]{\fnm{Shaifu} \sur{Gupta}}\email{shaifu.gupta@iitjammu.ac.in}

\author[3]{\fnm{Manoj Singh} \sur{Gaur}}\email{manoj.gaur@iitjammu.ac.in}

\affil[1,2]{\orgdiv{Department of Computer Science and Engineering}, \orgname{, Indian Institute of Technology Jammu}, \orgaddress{\street{Jagti}, \city{Jammu}, \postcode{181221}, \state{Jammu and Kashmir}, \country{India}}}

\affil[3]{\orgname{Indian Institute of Technology Jammu}, \orgaddress{\street{Jagti}, \city{Jammu}, \postcode{181221}, \state{Jammu and Kashmir}, \country{India}}}

\abstract{User profiling, the practice of collecting user information for personalized recommendations, has become widespread, driving progress in technology. However, this growth poses a threat to user privacy, as devices often collect sensitive data without their owners' awareness. This article aims to consolidate knowledge on user profiling, exploring various approaches and associated challenges. Through the lens of two companies sharing user data and an analysis of 18 popular Android applications in India across various categories, including \textit{Social, Education, Entertainment, Travel, Shopping and Others}, the article unveils privacy vulnerabilities. Further, the article propose an enhanced machine learning framework, employing decision trees and neural networks, that improves state-of-the-art classifiers in detecting personal information exposure. Leveraging the XAI (explainable artificial intelligence) algorithm LIME (Local Interpretable Model-agnostic Explanations), it enhances interpretability, crucial for reliably identifying sensitive data. Results demonstrate a noteworthy performance boost, achieving a $75.01\%$ accuracy with a reduced training time of $3.62$ seconds for neural networks. Concluding, the paper suggests research directions to strengthen digital security measures.}


\maketitle

\section{Introduction}
Personalized systems are designed to address the issue of information overload for users searching for specific queries. They achieve this by customizing information to individual users based on their profiles \cite{r2}. A user profile being a compilation of information associated with a user, can help to achieve this goal and to attain target of accurate recommendations and ultimately deliver personalized information.

As of year 2024, a vast array of data companies exists that track people across all aspects of their lives, both online or offline, collecting and accumulating an unprecedented volume of consumer data. This data provides insight into the behaviour and demographics of consumers and enable companies to analyze the patterns that can lift the overall customer experience. Consequently this data has become a valuable economic asset. For decades, regulatory authorities, journalists and civil society have mentioned the lack of transparent policies by the companies regarding buying and selling of their customer's data. 

In the data ecosystem, different companies continuously trade digital user profiles with one another. These data companies combine and link data from various devices of users such as computers, mobiles and diverse IoT devices. Every click on a website transmits information using invisible script embedded in web pages to hundreds of third-party companies. These invisible codes, called trackers, record information about a user visiting a website and collect data on how a user interacts with the visited pages. When the same tracker is present across numerous websites, it can develop a detailed profile of user's online activities and behaviour \cite{r7}. Similarly, every swipe on a smartphone may trigger these hidden data sharing mechanisms, collecting rich information about users. This information flow is not only limited to device manufacturers and application owners but also extends to a significant number of other third-party companies.

The increasing number of smart devices, has made the collection of personal data a greater threat to user’s privacy. In order to provide services through smartphones, applications require access to phone features such as the camera and microphone. However, many applications ask for permissions to access sensitive system resources (e.g., sensors like  microphone, camera and GPS), personal information of the user (e.g., email address and contact list), and also unique identifiers (e.g,. IMEI number) that are not required to provide services but to track users \cite{r157}. Vulnerabilities in the applications or operating systems of the devices can expose personal data. For instance, whatsApp vulnerability (CVE-2019-3568 \cite{r166}) allowed attackers to remotely install surveillance software on phones by calling the targeted device. This allowed the installation of spyware, compromising user's data on device, and potentially enabling further attacks. These kind of vulnerabilities creates a loophole for hackers and makes user's sensitive and personal information more vulnerable.

This study covers everything about how user profiling works and why it matters for privacy and security. Other studies usually focus on just one part—either user profiling process \cite{r2}, \cite{r12}, \cite{r14}, \cite{r100},  \cite{r101} or why it might be a problem for privacy \cite{r115},\cite{r118},\cite{r126},\cite{r127}. This research provides a complete repository that explains all about user profiling and talks about how it affects privacy and security.
Rest of the article follows the following structure. Section II covers objectives of this paper and Section III provides a review of literature in the domain of user profiling. Section IV describes user profiling and its use-cases. Section V illustrates the user profiling process while section VI provides study on two data brokers. Privacy and security concerns regarding user profiling are presented in Section VII. Section VIII describes an experimental study and implementation of framework to detect privacy leakage from a variety of Android applications. Section IX illustrates some open research areas. Article is concluded in Section X.

\section{Objectives}
This article conducts an extensive review of recent contributions exploring the latest practices and their implications. The contributions of this article can be considered in four parts as outlined below:

i)	Review of user profiling, including methods, types, models, and processes, along with highlighting challenges in current mechanisms.

ii)	Analysis of privacy and security issues present in the user profiling process. This study explores how thousands of commercial organizations collect, trade and make use of personal data, and influence the lives of billions of people. Based on the review of literature and articles, a study of two data collecting companies that operate by collecting, analyzing and selling user's data is presented. 

iii) An experimental study of several mobile applications used in India is covered to detect leaks of user's personal data by intercepting the network traffic. Further, ML classifiers that improve state-of-the-art frameworks in detecting personal information exposures is presented and then XAI algorithm LIME is used to provide explanations for the results generated by the classifiers.

iv) To the end, study provides several open research directions for future work.

\section{Related Work}
We briefly review some of the prominent approaches related to user profiling in this section, highlighting the existing shortcomings. We short-listed these papers based on relevance and clarity of understanding. Keyword search of ``user profiling'', ``privacy leak'', ``personal data leakage'' etc, assisted us in the short-listing process to identify relevance of different existing works to our text. 

Some studies have employed diverse platforms like social media and smart devices to deduce user actions, in order to develop user profiles and provide relevant recommendations. Table \ref{tbl1} provides summary of some of these studies. Table \ref{tbl2} presents a compilation of several studies aiming to address the security and privacy concerns associated with data collection for user profiling. These studies encompass various aspects, such as detecting potential disclosure of sensitive information from user devices and identifying embedded trackers in smart devices. Several other studies have focused on user profiling through surveys, and thus we present comprehensive information about these investigations.
 
\begin{table*}[!h]
\begin{scriptsize}
\caption{Summary of works in building user profiles}
\label{tbl1}
\resizebox{\textwidth}{!}{%
\begin{tabular}{p{0.5cm} p{2.5cm} p{2cm} p{7.5cm} p{4.5cm}}
\toprule
\textit{\textbf{Ref.}} & \textit{\textbf{Objective}} & \textit{\textbf{Data Source(s)}} & \textit{\textbf{Methodology}} & \textit{\textbf{Gap noted}}\\
\midrule
\cite{r114}& Build multidimensional profile of users& Twitter &  
\begin{itemize}
\begin{scriptsize} \vspace{-2.9ex}
    \item Developed model to represent dynamic connections \item Ranked list of users generated to find relevant information based on individual requirements.
\end{scriptsize}
\end{itemize}& Scale of the experiments or the diversity of used data is not explained. \\
\midrule
\cite{r101} & Develop framework to fuse diverse information from various sources.& - &  Used deep learning to predict user's traits to create profile. &\begin{itemize}  \begin{scriptsize} \vspace{-2.9ex} \item Used small datasets. \item Not considered privacy and ethical aspects of social media data. \end{scriptsize} \end{itemize} \\
\midrule
\cite{r100} &  Create smart-TV based recommendation system. & In-built resources of smart TV.&\begin{itemize}  \begin{scriptsize}  \vspace{-2.9ex} \item Detect faces in front of TV. \item Generate anonymous and consolidated user and group profiles. \item Create item profiles using stored videos, live channels, EPG data, and other sources. \item Recommend items to user by comparison between two types of profiles. \end{scriptsize} \end{itemize} & \begin{itemize}  \begin{scriptsize} \vspace{-2.9ex} \item Unsuitability for public places. \item Accuracy issue due to low brightness. \end{scriptsize} \end{itemize} \\
\bottomrule
\end{tabular}}
\end{scriptsize}
\end{table*}
\begin{table*}[!h]
\begin{scriptsize}
\caption{Summary of works on privacy and security aspects of data collection}
\label{tbl2}
\resizebox{\textwidth}{!}{%
\begin{tabular}{p{0.5cm} p{2.5cm} p{2cm} p{6cm} p{6cm}}
\toprule
\textit{\textbf{Ref.}} & \textit{\textbf{Objective}} & \textit{\textbf{Data Source(s)}} & \textit{\textbf{Methodology}} & \textit{\textbf{Gap noted}} \\
\hline
\cite{r115} & Identify clear text sensitive information from encrypted communications. & Medical IoT devices&Isolated traffic originating from fixed set of IP addresses& Differentiation of encrypted traffic from compressed clear-text traffic not done.\\
\hline
\cite{r116} & Identify smartphone based on data transmitted via its sensors. &  Sensor of smartphones & Developed a web page to collect sensor data & \begin{itemize}  \begin{scriptsize} \vspace{-2.9ex} \item Gyroscope lacks manual calibration   \item Complexity of real-world motion sensor tracking not studied. \item Fails to address practical constraints of implementing proposed defences. \end{scriptsize} \end{itemize}\\
\hline
\cite{r117} & Show encryption alone is insufficient to preserve privacy of smart homes & IoT devices & Separate traffic into packet streams and label it by type of device and then correlating traffic rates with user interactions to infer consumer behaviour.& \begin{itemize}  \begin{scriptsize} \vspace{-2.9ex} \item Only focused on passive network threat model \item Identify same manufacturer device using DNS is limitation. \end{scriptsize} \end{itemize}\\
\hline
\cite{r118} & Check presence of trackers in OTT devices. &  OTT TV streaming devices & Built crawler to interact with OTT channels. & \begin{itemize}  \begin{scriptsize} \vspace{-2.9ex} \item Only focus on Roku Express and Amazon Fire TV Stick \item Crawler has restricted capability. \end{scriptsize} \end{itemize} \\
\hline
\cite{r127} & Identifying PII or ad requests in HTTP packets. & Android apps. & Used Federated learning to classify outgoing HTTP packets. & \begin{itemize}  \begin{scriptsize} \vspace{-2.9ex} \item Not handled data and system heterogeneity. \item Didn't address real-device resource constraint problem \item Didn't consider possibility of attackers among clients in a sub-network \end{scriptsize} \end{itemize}\\
\hline
\cite{r126} & Reveal PII leakage and give users control over their data. & Popular apps on iOS, Android, and Windows.& Automated the disclosure of PII leakage using machine learning on a cross-platform scale. & \begin{itemize}  \begin{scriptsize} \vspace{-2.9ex} \item Used manual labeling and crowd-sourcing from a specific group of users. \item Relied on a centralized server model, raising privacy concerns \end{scriptsize} \end{itemize}\\
\bottomrule
\end{tabular}}
\end{scriptsize}
\end{table*}

Stewart et al. \cite{r132}, conducted a survey on methods for filtering vast amounts of information from electric sources, including the internet, to create user profiles. These techniques includes statistical term-based, neural networks and social filtering. \cite{r135} discussed user modelling techniques for constructing and representing profiles for social media platforms, highlighting their strengths and weaknesses and providing a vision for future research, for example, creation of more dynamic and intelligent profiles to receive more appropriate results from users’ profiles. \cite{r15} reviewed the latest advancements in user profiling, including methods, features, and taxonomies. It explored techniques such as data acquisition, feature extraction, and profiling approaches, along with performance metrics. The survey also addressed challenges like dataset size, the cold start problem, and domain dependencies. Furthermore, it outlined future research areas, such as developing versatile, dynamic, language-independent user profiles.

\cite{r133} reviewed studies related to profiling smartphone users through ordinary apps, presenting a general framework for learning user information from smartphone applications. They included the method of data collection, pre-possessing, and user profiling, with implications and suggestions for improving business services, user experience, and profits, and developing mobile context-aware tools to improve the quality of life of users in different aspects. Another work by \cite{r134} provided an overview of profiling users on social networks.
Since the data available on the web ranges from semi-structured to unstructured, various approaches were presented to profile users in online social networks. These approaches included clustering, face detection, user activities, content analysis and behavioral analysis. Authors \cite{r136} conducted a systematic mapping study of profiling users based on reviews, presenting the latest trends in user profile modeling and analysis. 

\cite{r131} examined intrusion detection and prevention systems from the standpoint of exploiting behavior including \textit{system behaviours} that is generated by hosts and networks and relate to the host activities and network status and \textit{user behaviours} that relate to the direct interaction between the user and the system, for example, typing patterns. These behaviors were examined to determine whether a user was legitimate to be on the system i.e. review of intrusion detection and prevention systems for profiling users. In the first step of profiling users, the behavior was analyzed, then categorised into system behaviors. As a result of this classification, data profiles were then divided into system profiles and user profiles, with the latter being further categorized into more specific categories namely biometric and psychometric profiles based on their characteristics. A summary was then provided of the advantages and limitations of these specific profiles and related analysis techniques.

In contrast to these studies, our work presents an examination of the user profiling process pipeline reviewing latest work, starting with deciding the context of profiling users, collecting data from various platforms using numerous approaches, and finally constructing the profile using different techniques. This is followed by security and privacy implications that profiling has on user's life. Also, a case-study on two companies has been carried out which are involved in profiling process and establishing links between a profile to an individual on web. The identification of privacy leakage from android apps justifies the data exfiltration process that occurs from user’s devices without their prior consent and knowledge. Next we discuss in detail user profiling process along with its usecases in Section V.

\section{User Profiling}
User profiling refers to the process of extracting data about the interests, preferences, behaviors, and needs of a user, which can be used to identify an individual. Various interpretations of user profiling have been proposed in the literature, depending on the context in which it is used. For instance, in work by \cite{r13}, user profiles are designed on a mobile cloud environment to provide distributed IT services and resources to users based on context information. Authors of this work manifests user profile as information of user and service consumed by them. Information like user ID, user name, hobbies, personal desire, and other details are stored in user's profile, while the service information part stored data about IT services used,  such as service name, context, provider, and frequency of access, etc. 

\subsection{The importance of User Profiling}
User profile is a fundamental concept in personalisation system. These systems customize content based on individual behaviour, interests, and preferences. This is used to enhance user's experience over the web and determine their intention \cite{r15}. In modern digital era, personalization is vital, and profiles form the basis of advanced technologies used to provide benefits to users. For example, insurers in the United States and the United Kingdom are pushing for wider acceptance of ``telematics'' devices in automobiles to gather real-time reports on a driver's behavior, which can be used to determine insurance costs \cite{r74}.

This way understanding user needs is vital for businesses to deliver tailored services. Personalized systems filter information to match user interests and adapt to different contexts. The next section covers different areas where user profiling has been used for years.
\subsubsection{Monetarily-driven entities}
Online behavioral advertising, also known as interest-based advertising, involves collecting data like search history or app usage, from a user's device to understand their behavior and preferences. This information is then used to deliver targeted advertisements to users. Many companies, including Google and Facebook, heavily rely on this practice for revenue generation. In 2023, Google, Facebook, and Amazon projected advertising revenues of 39\%, 18\%, and 7\%, respectively \cite{r113}. Amazon serves as a notable example of effective user profiling. In just two years, Amazon's revenue  generated through advertising sales has gone from 11 billion in 2020 to 31.16 billion U.S. dollars in 2021 and is forecasted to reach  64.3 by 2026 \cite{r122}. The company collects data from user interactions on its site, considering factors like time spent on pages, items added to the cart, and purchases made. Amazon also incorporates external datasets, such as census data, for demographic information. Amazon Privacy Notice, explains that the company collects information such as personal details, browsing information, and device specifics, allowing Amazon to create comprehensive customer profiles. Company employs collaborative filtering in its recommendation technology, suggesting products based on the preferences of customers with similar profiles. Additionally, Amazon extends its advertisements to other platforms, creating a holistic view of customers to enhance their personalized experience.

\subsubsection{Non-monetary-driven entities}
The public sector also utilizes user profiling to personalize e-government services offered to citizens. Personalized portals benefit citizens with services they specifically require, thereby increasing satisfaction levels. Profiling also aids in efficient and effective communication, deducing and anticipating citizens’ behavior and manipulating them to target specific section of society for a general cause like women health care hygiene. This gives governmental sector organizations enormous capabilities for their e-government strategies \cite{r59}. Apart from providing e-services, the government makes use of user profiling by collecting citizens data for various affairs, such as predicting crimes and illicit activities based on citizens' behavioural profiles developed over time. Also, there are organizations that use internal systems to track and recommend services for its employees based on their profile. For instance, IBM \cite{r158} utilizes data analytics and AI algorithms to match employees' profile with relevant projects and opportunities. Accurately representing the user profile is crucial to obtain the best results that closely align with their preferences. The next section will explain the details of the user profiling process.

\section{User Profiling Process}
\begin{figure*}[!h]
    \centering
    \scalebox{1}{
    \rotatebox{90}{\includegraphics[width=1.2\textwidth,height=0.50\textheight]{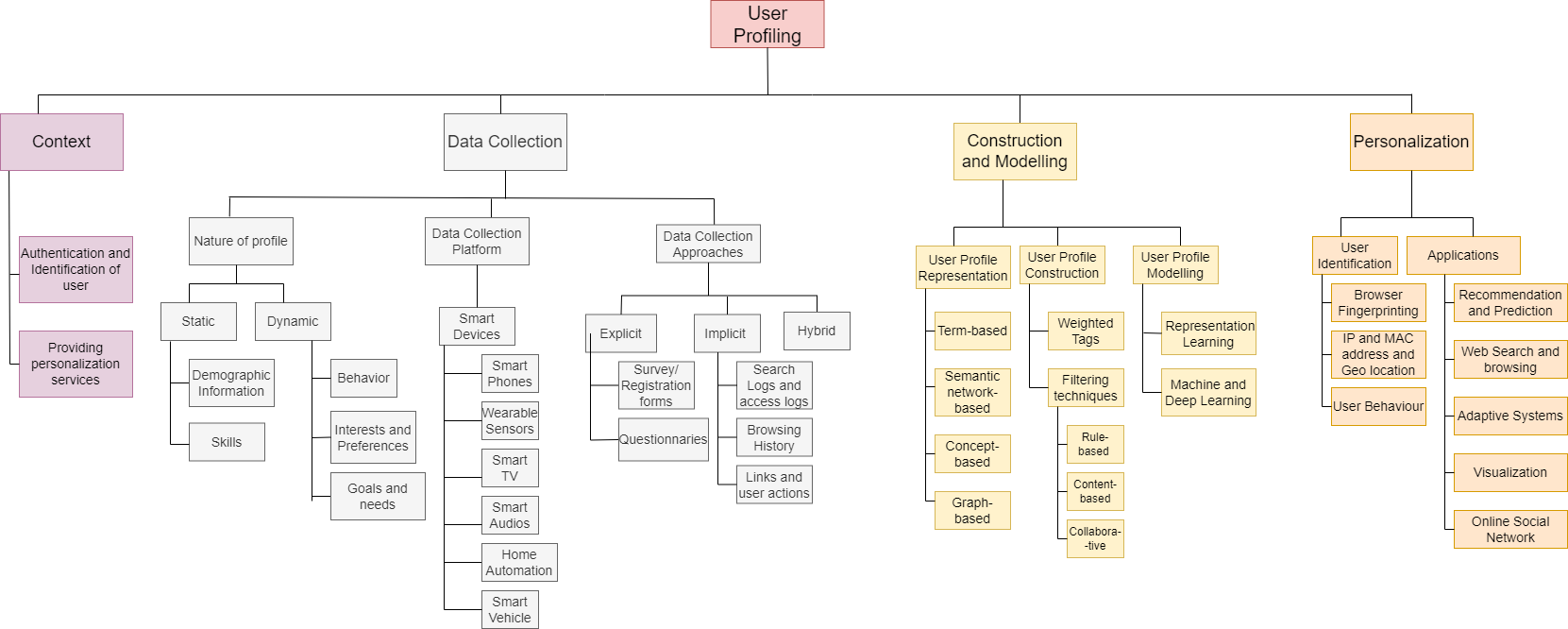}}}
    \caption{\label{figure2}Taxonomy of user profiling}
\end{figure*}

Numerous factors contribute to the creation and usage of a user's profile. This section covers a study of how a user is profiled. The taxonomy in the form of different phases in profiling process is depicted in Figure \ref{figure2}. The process of building a user profile starts with the data collection process, which consist of selecting the context of building the user profile and then selecting the content for creation of that profile. Raw data about the user is gathered from various platforms using various methods. The second step constitutes building and maintaining of user profile model, which consist of ways to represent a user model and various techniques that can be adopted for its construction and modelling. In the third step, personalized services are provided, which starts with identifying user on web and then matching their characteristics to the content of the profile. Finally, customized services are delivered to the matched user. Following sections contain a detailed description of every step of this process.
\subsection{Data Collection}
\subsubsection{Nature of profile}
Different types of data can exist in a profile ranging from user’s demographic data, skills, needs and goals, behavior, interests and preferences. Depending on the nature of data constituting the profile, user profile can be either static or dynamic. \textit{Static profile} is the one that contains user information that does not change or get modified. A work, \cite{r16} explained static profiling as a method to evaluate user’s static and predictable characteristics. This profile keeps user information for a longer duration. Manual information gathering methods are used by the static profiling that requires user's intervention to collect and analyze his static and predictable characteristics. \textit{Dynamic profile} in contrary to the static profile is automatically created by the agents or models deployed in the system and therefore, the user attributes and contents get updated with time. 

\subsubsection{Data Collection Platforms}
Gadgets like phones, watches, TVs, speakers, refrigerators, air conditioners, cars, etc have now become the most important source of personal data. 
In addition to providing a service to their users, these devices collect personal information via diverse applications and model user behavior. As an example, \cite{r23} employed the gyroscope and accelerometer sensors of smartphones to gather user data and create a behavioral model. The authors introduced a context-generation method that continuously updates a trust score, indicating the likelihood of users engaging in activities similar to their regular patterns. Whereas, \cite{r24} used smart watch and smart phone to get details of user’s activity and accelerometers sensor data to categorize user behaviour.

Web is an important source of data. Within seconds of a user clicking on a website or interacting with a mobile application, trackers embedded in website/app profile a user. One usually needs to create an account in order to use services of most web platforms. User personal information such as their name, email id, birthday, profile picture, and possibly even the financial information are all linked to his account. All the interactions a user has while using social media service and e-marketplace, including the type of content or product user finds interesting and shared by him, interactions with other users and advertisements the user interacts with and responds to are identified with this account. A profile can be developed based on the data a user provides, and also their interactions. \cite{r114} and \cite{r18} created a user profile from the personal information that is shared by users on social network sites. Websites can also study their users' shopping habits, how did they find their website, and if they are interested in advertisements on pages they visit. With help of IP addresses, cookies, and small image files called tracking pixels or web beacons, most of this information can be reverse engineered. By combining information from different websites, a detailed profile of a user can be generated. For example, \cite{r17} and \cite{r3} generated user profiles from data gathered from user's web search with an aim to get insights into user behavior patterns.

Data collected from various platforms can be used by businesses to gain a deep understanding of their customers and provide them with personalized services.

\subsubsection{Data Collection Approaches}
The collection of user's personal data from diverse platforms mentioned above has always been a challenge, and since huge data is gathered and the most important ones have to be identified and filtered. The collection of data from user can be achieved through explicit, implicit, or hybrid approaches. \textit{Explicit} or \textit{Opt-in} data collection methods use manual techniques that require user intervention, such as personalized applications that ask users to provide ratings and feedback, or surveys and registration processes used when registering for a service on a website \cite{r25}. Profiles constructed using an explicit approach are static and only accurate until the user changes their interests and preferences. The \textit{Implicit} or \textit{Observed} data collection approach involves the use of automatic agents or models to collect data without user intervention. Farid et al. \cite{r26} proposed an agent-based model that builds an implicit user profile from bookmarks and browsing activities, utilizing web and search logs. This approach aims to analyze a user's interests by processing search records, access records, and browsing histories, and recording user actions such as tagging, bookmarking, and downloading while browsing web pages. \textit{Hybrid} or \textit{Inferred} approach takes advantage of both implicit and explicit techniques to provide accurate recommendations and services. Hybrid approaches result in profiles that are more adequate and accurate as they periodically update user information. Fakhfakh et al. \cite{r28} utilized a hybrid approach for collecting user preferences. They consulted users for their preferences using the explicit approach, while the implicit approach discovers user preferences through MovieLens datasets, wherein a mapping between MovieLens genres and user preferences is established based on their semantic meanings and relationship. Next, we will expound upon the methodology of formulating and modelling the user profile utilizing diverse techniques.

\subsection{User Profile Construction and Modelling}

After completing the data collection phase, the subsequent step involves choosing a suitable method to represent user profiles, then creating profiles employing diverse methodologies followed with development of a model aimed at predicting user preferences. Detailed descriptions of these processes are provided below.

\subsubsection{User Profile Representation}
The efficacy and accuracy of user profiling also relies on the appropriate representation of collected information. In this section, we will discuss the various types of user models based on their representation.
\begin{enumerate}
\item Term-Based (vector-space model): In vector-based model, each profile is composed of a vector of keywords, where each keyword represents a topic of interest to the user, and associated weights that represent the numerical representation of the user's interests related to those keywords. When a search query is submitted, the system retrieves hypermedia documents present in web that are converted into weighted keyword vectors. The Antagonomy system by \cite{r32}, for instance, used this model to create a personalized online newspaper that learned personal preferences from user behaviors. However, one major drawback of the keyword-based profile is polysemy ambiguity, which means keywords may have multiple meanings. For example, the word ``foot" can refer to either a body part or a scale unit, which may lead to inaccuracies in the user profile \cite{r26}.
\item Semantic Network-Based (ontology-based): To address the issue of ambiguity due to polysemy in keyword-based profiles, weighted semantic networks are used to create profiles where each node represents a concept. Initially, the network comprises a set of disconnected nodes. As additional user information is gathered, the profile is enhanced by incorporating more keywords related to various concepts. Subsequently, links are introduced to depict the connections between these concepts. This method offers a way to represent profiles based on descriptive keywords and knowledge using a consistent system that has been developed over a number of years. One such system is WordNet, which is an extensive lexical database of the English language. It stores a knowledge base that organizes English words into semantic relations known as synonym sets. Users and their semantic preferences are represented by the information inherent in an existing ontology. The InfoWeb \cite{r34} system, which filters digital library documents online, utilizes semantic networks to build profiles containing the long-term interests of users.
\item Concept–Based (hierarchy-based): Similar to semantic network based approach, in this approach too user's interests are represented in a form of a graph. However, here nodes represent abstract topics of interest to the user. Profiles are represented as vectors of weighted features, where features represent concepts instead of individual words or phrases. \cite{r36} proposed the use of hierarchical concepts, which allows for generalization. The hierarchy of concepts can either be fixed or dynamic, depending on the user's interest.
\item Graph-Based: Graph-based representations, however, focus on utilizing graphs or networks to represent and analyze user-related information without specific emphasis on semantic relations or abstract concepts. Nodes in these graphs signify entities or concepts, while edges represent relationships or connections between them. This approach captures intricate relationships, dependencies, and interactions within a user's profile. For instance, Chen et al. \cite{graphbased} proposed a method for enhancing user profiling within sequential recommender systems using two layers of graphs: a global graph to record transitions among various behaviors across all users, and a personalized graph to model individual user interactions and preferences.
\end{enumerate}
\subsubsection{User Profile Construction}
Numerous techniques exist that can be adopted for constructing user profiles, contingent on the context. The resultant profiles should ideally possess a dynamic disposition, that is, they must be adaptable enough to accommodate any alterations in the user's immediate or long-term interests, and should be updated periodically. The ideal approach is to opt for a construction technique that requires minimal user intervention and feedback. A few of the construction methods reported in the literature are discussed below:

\begin{enumerate}
    \item Weighted Tags: One significant strategy for constructing a user profile is to assign weights to the terms. Different techniques, such as Boolean or Frequency weighting, are used for this purpose. The underlying principle behind these techniques is that the occurrence of terms within documents characterizes the user profile to which the document belongs. Thus, the more often a term appears in a class, the more it reflects the characteristics of that class. Similarly, frequent usage of certain tags mean higher user interest in associated topics \cite{r121}.
    \item Filtering techniques: In this approach, user profiling makes use of filtering mechanism for retrieving pertinent information that aligns with the user's specific requirements. Various filtering techniques that can be employed for this purpose are given as:
    \begin{itemize}
\item Rule-based approach: This approach involves selecting pertinent user information using a predefined set of ``if-then" rules. Pre-defined categories of users are defined to determine which content can be incorporated into a user's profile. For instance, online brokerages usually classify their accounts based on gender and age and then offer distinct services, products, or benefits based on these categories. However, the challenge with this approach is that it is difficult to obtain marketing rules from domain experts to validate the effectiveness of the extracted rules \cite{r14}.

\item Content-based filtering: Content-based filtering is an approach for creating user profiles by comparing item profiles with user data. This technique identifies the content of the item by extracting keywords from the product descriptions. These are more commonly used in text-intensive domains \cite{r11,r14}. Also, analyzing limited content can be challenging and may reduce the performance of the approach, particularly in domains that involve multimedia content like images and audio.

\item Collaborative filtering: This approach operates on the idea that similar users will have similar preferences. Therefore, this method clusters similar interest users into groups. It uses an algorithm that aggregates feedback from multiple users to suggest items that will appeal to target users by considering similarities between different users \cite{r124}. These algorithms fall into two categories: memory-based and model-based. Memory-based methods, such as correlation analysis, compare the active user's profile with similar ones in the database, relying on the entire user-item matrix to tailor recommendations. Model-based methods learn from the entire dataset to offer personalized suggestions without needing the entire dataset in memory during runtime, providing scalability advantages. For example, Amazon.com utilizes collaborative filtering to personalize web pages with tailored recommendations based on individual customer interests.
\end{itemize}


\end{enumerate}

\subsubsection{User Profile Modelling}
After constructing users' profiles, the subsequent steps entail developing a computational model capable of predicting user needs and preferences. Below are several approaches used for modeling users.
\begin{enumerate}

\item Using representation learning: This approach emphasizes modeling users by learning latent representations for each user through the utilization of items, item features, and/or user-item response matrices. Here, items refer to any entities that users interact with, such as products in an e-commerce platform, articles in a news website, or movies in a streaming service. They represent the objects of interest for users. Representation learning enables the extraction of meaningful latent features from both static data (e.g., tabular data) and sequential data (e.g., time-series data). Li et al. \cite{repres} conducted an extensive analysis of recent advancements in user modeling, with a specific emphasis on representation learning techniques. The study categorizes these techniques into two main types: static and sequential representation learning. Static learning methods, such as matrix factorization and deep collaborative filtering, are employed to capture user preferences and item characteristics within a static context. These methods are fundamental to numerous recommender systems and play a vital role in comprehending user behavior. The study also discusses sequential learning methods, including recurrent neural networks, which are designed to capture the dynamic evolution of user preferences over time.
\begin{table}[!h]
\begin{minipage}[t]{0.48\linewidth}
\centering
\caption{Survey of works using Machine learning techniques}
\begin{scriptsize}
 \begin{tabular}{p{0.75cm}p{2cm}p{3.25cm}}\toprule
      \textbf{Ref.}& \textbf{Model} & \textbf{Objective} \\ \hline
     \cite{ml1} & Random forest and neural network & High-risk user identification model\\ \hline
     \cite{ml2}, \cite{ml3} & Boosting algorithms, Naïve Bayes & Student profile modeling \\ \hline
      \cite{ml4} & k-means clustering  & Recommendation model\\ \hline
    \cite{ml5} & PCA & Anomaly detection \\ \bottomrule
    \end{tabular}
\end{scriptsize} 
\label{machinelearningtable}
\end{minipage}
\hfill
\begin{minipage}[t]{0.48\linewidth}
\centering
\caption{Survey of works using Deep learning techniques}
\begin{scriptsize}
\begin{tabular}{p{0.75cm}p{2cm}p{3.5cm}}\toprule
      \textbf{Ref.} & \textbf{Model} & \textbf{Objective} \\ \hline
     \cite{dl1}, \cite{dl2} & Attention networks, graph attention networks & Modeling user behavior. \\ \hline
     \cite{dl3}, \cite{dl4} & Neural network & Modeling user preferences. \\ \hline
      \cite{dl5}, \cite{dl6} & Convolutional neural network (CNN) & Modeling user representations and interests.\\ \hline
    \cite{dl7}, \cite{dl8}, \cite{dl9} & Auto-encoders & Analyzing and modeling user behavior and representations. \\\hline 
    \cite{dl10}, \cite{dl11}, \cite{dl12} & Recurrent neural network (RNN) & Personalized recommendations, student performance prediction, and user behavior modeling.\\\hline 
    \cite{dl13}, \cite{dl14} & Transformers & Generating recommendations and modeling user behavior.\\
    \bottomrule
    \end{tabular}
\end{scriptsize}
\label{deeplearningtable}
\end{minipage}
\end{table}
\item Using machine learning: One way to create user profiles is through the use of machine learning methods, where the aim is to predict users' preferences based on a small number of instances of data about users and their behavior. An example of this is the k-NN algorithm, which can be used to classify each user's preferences from a collection of labelled users' preferences. A summary of works using machine learning techniques is presented in Table \ref{machinelearningtable}. \cite{ml1} analyzed user data from product discussions, employing a supervised random forest model to identify high-risk users who have posted negative comments in new public opinions, supplemented by a backpropagation neural network. \cite{ml2} investigated boosting algorithms for student profile modeling, exploring how student behaviors and traits influence academic performance through adaptive learning support. \cite{ml3} focused on student performance prediction, comparing Support Vector Machine (SVM) and Naïve Bayes classifiers. \cite{ml4} applied partitional clustering algorithms, particularly k-means, to group profiles based on similar interests and preferences. To improve accuracy of recommendations in e-commerce environment, \cite{ml5} utilized principal component analysis (PCA) to extract important features and then build anomaly detection system to examine user behavior in database systems and web browsing environments.
Deep learning has emerged as a prominent approach for uncovering intricate patterns within user data. Table \ref{deeplearningtable} offers an overview of research employing deep learning for user behavior modeling. In \cite{dl1}, user behavior was modeled by exploiting semantic similarities between the geographical proximity of items and user requests. Meanwhile, \cite{dl2} focused on predicting user identity links across social networks to improve recommendations using multi-layer perceptron. Their approach utilized a heterogeneous graph representation and multiple attention layers for aggregating user information.

In a different study, \cite{dl3} proposed a two-stage collaborative filtering recommendation model. This model incorporated a time-aware attention mechanism for dynamic user preferences and a matching function learning model based on deep matrix factorization and multiple-layer perceptron for user-item feature interactions. In personalized recommendations, \cite{dl4} aimed to enhance user profiling for point-of-interest recommendations using graph neural network and attention mechanisms. Their architecture included layers for encoding inputs, aggregating neighborhood entities, and integrating user/location representations.

\cite{dl5} introduced a temporal CNN approach for continual user representation learning across tasks, leveraging partial parameters from previous tasks, while \cite{dl6} developed a candidate-aware user modeling framework for personalized news recommendation, using
CNN networks to model local click behavior context.

In other domains, \cite{dl7} employed stacked autoencoders and clustering for user behavior analysis for power grid user behavior analysis, and \cite{dl8} employed a variational autoencoder-based model for user representation learning using end-to-end learning algorithm. For user modeling in inconsistent client environments, \cite{dl9} proposed a federated learning method.

\cite{dl10} introduced a user-based RNN model for personalized recommendations. In student performance modeling, \cite{dl11} developed a personalized federated learning framework and proposed user profiling using semantic behavior modeling  utilizing attention-based Gated Recurrent Units (GRU). \cite{dl12}, proposed user profiling by combining semantic behavior modeling with RNNs.

For user preference modeling, \cite{dl13} utilized a Transformer-based architecture with self-attention mechanisms  for contextualized item embeddings. While \cite{dl14} introduced UserBERT for pre-training user models on unlabeled data, employing contrastive self-supervision and behavior sequence matching tasks.
\end{enumerate}

Once the user profiles have been constructed, they can be leveraged to provide personalized services to users. The next section will explore two prominent data brokers that profiles users and then sell their profiles to its clients. 

\section{Data brokers: Understanding the role and practices}

A data broker is a business that gathers a large amount of personal information about people from various online and offline sources and sells it to clients. This information is used for identity verification, credit assessments, employment decisions, insurance, housing, and marketing products. Brokers like Acxiom  \cite{r50} also purchase data from variety of other sources, analyze it, and refine it for future marketing purposes. Marketers identify what they can buy from these large data brokers because this data provides insight into their current audience. Additionally, this data can assist in the development of strategies aimed at expanding a company's customer base \cite{r43}. 

Data management platforms (DMPs) help brokers integrate customer data with profiles from third-party sources. The platforms synchronize with each other, for instance by employing cookie synching, to recognize the user across various tracking entities and match identifiers from disparate systems. DMPs have capacity to recognize users not only on the Web, but also across different devices, by anonymously matching data from multiple sources \cite{r41} using identifiers discussed in Section 6.3. 
These identifiers are then transformed into an alphanumeric string via cryptographic methods. Thus, a user can be recognized as the same individual across various devices and platforms, and their profile can be connected to more comprehensive data.

Today, users use multiple platforms and devices, making it challenging for data companies to match user data accurately. \textit{Deterministic matching} and \textit{probabilistic matching} are two main approaches used for this purpose, with deterministic matching relying on common identifiers like email addresses or mobile identifiers, and probabilistic matching using algorithms and statistical models based on various data pieces. After review of existing literature and going through articles, details on two data companies are presented for examination and analysis that collect user's data from diverse domains and concatenate them into profiles for sharing it further.

\subsection{Acxiom}
Acxiom is one of the major players in the field of consumer data brokering and collection, claiming on their website that their primary focus is on providing the necessary data, technology, and services to enhance customer experiences worldwide \cite{r50}. Acxiom Marketing Solutions (AMS) functions as Acxiom's data brokerage, while LiveRamp is a Software-as-a-Service (SaaS) platform that helps marketers utilize their existing data sets. AMS gathers information from a variety of sources, including publicly accessible records (such as data on births, deaths, marriages, divorces, and changes of name and address), website cookies, surveys, and other digital tracking services. Data is combined and sorted into various user categories. Sorting in this way goes beyond simple demographics. For example, Bruce Schneier, a privacy expert, noted in his book ``Data and Goliath" that Acxiom categorizes its data collection into categories like ``potential inheritor," ``adult with senior parent" and ``households with a diabetes focus or seniors needs". These lists can then be sold to companies looking to market to these specific groups  \cite{r77}. LiveRamp's primary service is ``data onboarding" or ``offline matching," which involves linking offline user records to online identifiers tied to a browser or device. This enables LiveRamp's clients to target users based on various behavioral and attitudinal data, such as purchase history, TV show preferences, website traffic, and smart device usage \cite{r78}. According to LiveRamp's Service Policy \cite{r78}, this approach allows advertising firms to display targeted ads to specific individuals across email, mobile apps, websites, and addressable TVs.

\textit{Unique Identity Resolution:} Acxiom's AbiliTec is a highly effective recognition solution that provides clients with a unique identity resolution service. By using various offline and online touchpoints and multiple variables, AbiliTec goes beyond the basic name and address information to achieve more precise recognition of individuals and households. Acxiom's Data Services API \cite{r47} states that the AbiliTec Link not only connects consumer records across different databases, platforms, and devices, but also assigns a single link for each record, allowing for a complete view of the user. As a result, individuals and their families can be identified instantly using the ``hashed entity representation", which utilizes phone numbers, email addresses, smartphone IDs, or any combination of these, as well as name, address, city, and zip code.

\subsection{Oracle}
One of the largest multinational technology companies, Oracle, has been acquiring several data brokers and consolidating their operations under the umbrella of Oracle Data Cloud \cite{r161}. By merging information from many other data brokers, Oracle has amassed an extensive repository of consumer data, which it has integrated into a full service digital marketing platform \cite{r81}.

To enhance targeted advertising, Oracle integrates offline data from loyalty card programs, which are used by retail businesses to gather customer information, with digital media. This connection between offline behavior and online profiles enables a deeper understanding of consumers. Online profiles are generated using cookies saved on a consumer's device from partner websites. With Oracle Data Cloud, consumer interests are tracked across both online and offline activities and displayed to advertisers. This allows advertisers to utilize first-party and third-party data to create personalized marketing campaigns and target specific audiences. Additionally, Oracle provides data and campaign metrics in one convenient location. By using first-party data, easily accessible third-party data, and new second-party data, companies can build comprehensive customer profiles with Oracle. Oracle's website and privacy policy indicate that personal information may be shared with third-party companies for commercial purposes. As per Oracle's privacy policy, user information is collected through various methods. In certain instances, data is directly gathered using email addresses provided by users during interactions with Oracle or its partners. In other cases, unique identifiers like mobile device identifiers or cookie IDs on browsers are used to gather user data. This information can be associated with interest segments or profiles. Interest segments refer to a group of users who share similar preferences or behavior and are utilized for direct marketing by Oracle's customers. Profiles comprise a collection of attributes about a specific user or device, or a group of users or devices that share similar attributes, which are used for marketing by Oracle's clients.

\textit{A unique ID and matching process for Consumers:}
After generating data-driven audiences, a business must establish connections with them, which can be challenging due to single user using multiple devices. Each browser has its own set of cookies, while on mobile devices, all applications share a mobile advertisement ID that functions similarly to a cookie.

Oracle ID Graph (OIDG) connects these ID sources and validates them with accuracy against a high-quality data that is known to be true because it is made up of verified transaction and subscription data \cite{r90}. OIDG consolidates all interactions across various channels in order to develop a single actionable user profile. To identify users on both online and offline platforms, all of Oracle's clients send their match keys (which can be any unique user ID) to the company. An encrypted email is the most common type of match key, as it can be gathered either online or offline. In addition to Oracle hashed IDs (normalized SHA-256 hashed email address or phone number), clients can utilize encrypted or hashed unique user IDs (UUIDs) based on phone numbers, email addresses, physical addresses, client account numbers, and even IP addresses. Oracle synchronizes match keys with the network of user profiles that are connected by OIDG. All Oracle clients use OIDG to manage their IDs and attributes.

The data transmitted by various applications is gathered not only by organization that owns that applications (for example, Facebook is owned by Meta Platforms, Inc.) but also by various other data companies that work mainly by gathering, tracking and linking a user to a profile and then selling further to its clients.

\subsection{Tracking and linking user profiles}
\begin{figure*}[!ht]
\centering
\includegraphics[width=0.95\linewidth,height=0.20\textheight]{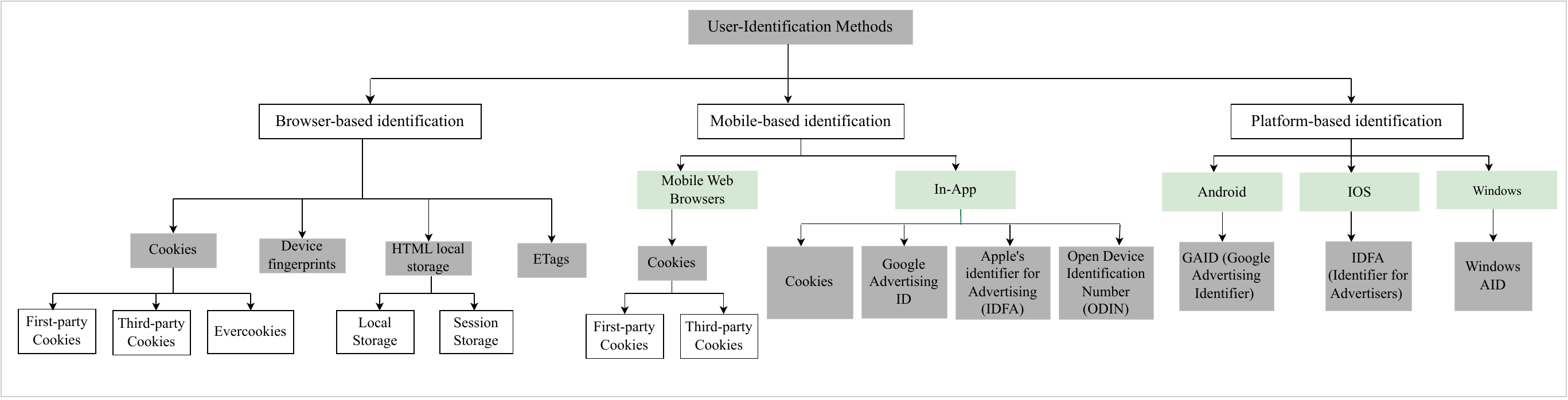}
\caption{\label{figure-2}Identifiers used to track users across devices and platforms}
\end{figure*}

In Figure \ref{figure-2}, an overview of the identifiers used for tracking users across various browsers, devices, and platforms is presented. The ability to identify users is primarily determined by the specific device being used and whether they are accessing content through a web browser or a mobile app.

Cookies, device fingerprints, HTML local Storage, and ETags are commonly employed in \textit{browser-based }identification methods.
\begin{itemize}
\item Cookies are small files that websites send to a user's browser. These files enable tracking and monitoring of the websites visited by the user and the specific items they interact with or click on during their browsing sessions. ``First-party" cookies, are generated by the domain visited by a user, and ``third-party" cookies, created by domains other than the one being visited, represent the two types of cookies. Third-party cookies allow advertising companies to monitor users' browsing activities. Another variation of cookie is ``ever cookie", a persistent type of cookie that is stored in multiple locations within the user's browser and device. Even if deleted, an ever cookie has the ability to restore itself from alternative locations.
\item Device fingerprinting is the practice of recognizing and monitoring individuals by analyzing the unique attributes of their devices. This involves collecting various details like browser version, operating system, language, installed plugins, and settings to establish a distinctive identifier.
\item HTML5 Local Storage provides a means to store and gather user information. There are two types of local storage: ``local storage" for data retention without an expiration date, and ``session storage" for temporary data during a specific session. In comparison to cookies, HTML5 local storage offers superior capacity and accessibility, while eliminating the need for web server call \cite{r48}.
\end{itemize}
For \textit{mobile-based} identification, identifiers used to track users are:
\begin{itemize}
\item Cookies as explained in browser-based identification methods, are also used in case of Mobile-browsers.
\item For mobile applications, identifiers used are Google's advertising ID (which is a 32-digit string of characters), Apple's IDFA and Open Device Identification Number (deprecated). 
\end{itemize}
The last classification is \textit{platform-based} i.e., based on type of platform being used by the user:
\begin{itemize}
\item Android uses Google's advertising ID. For example, the advertising ID of one of our android test devices is ``de7d3063-968a-4450-932c-7abf87a0a261" which can be accessed in the device settings under Google $\longrightarrow$ Ads.
\item iOS uses Apple's IDFA. For one of our iOS testing device, identifier is ``3F946465-AE34-4E7E-AF7E-C3C4A4647CCA" and can be acessed using an app named ``Get My IDFA" \cite{r162}.
\item Windows has its own advertising ID.
\end{itemize}

\section{Privacy and Security issues of user profiling}
While user profiling can bring benefits to various domains, it also poses potential security and privacy risks. In the realm of online platforms, data privacy and security are interrelated concerns. Data privacy concerns the authorized access of user data, while data security focuses on implementing measures to protect privacy in the event of any violation.

Online activities such as searches, browsing histories, purchases, and social media engagement can provide digital cues about an individual's interests and personality. For example, IBM's Personality Insights service can create a detailed user profile from digital communication, including text messages, emails, tweets, and forums \cite{r73}. This detailing goes beyond just demographics and location data. Online behavior can reveal various aspects of a user's personality, such as their level of extroversion, environmental awareness, political inclination, or travel preferences. Companies collecting large amounts of data from both online and offline sources can use this information to create rich user profiles for billions of people.


The hardest part of living in a digital world is in striking a balance between free service consumption and the disclosure of personal information to businesses for sale. Platforms such as Google aim to streamline users' digital and physical lives, offering convenient and cost-effective services. These platforms collect user data to improve their services and enhance customer satisfaction. Since users prefer such platforms, they willingly provide their personal information, thereby facilitating data collection by these platforms.

\subsection{Privacy Issues}
\begin{figure}[h]
    \centering    \includegraphics[width=0.5\textwidth,height=0.40\textheight]{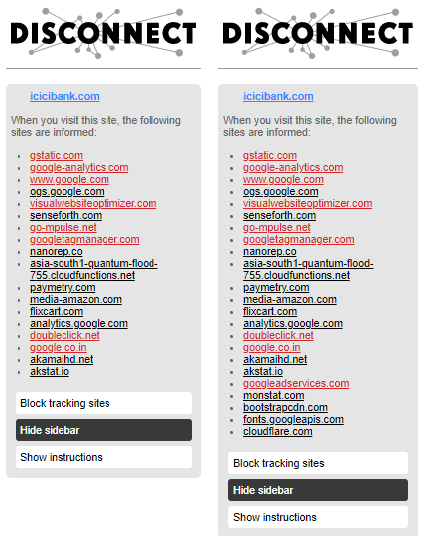}
    \caption{\label{figure3}Source: Disconnect.me. This figure illustrates the presence of trackers on the ICICI bank website. Colored domains signify tracking sites, while gray domains may also track users. Left part: domains informed when the user opens the website without logging in. Right part: domains informed when the user has logged in. (Accessed Date: October, 2021)}
\end{figure}
Data-driven companies can derive substantial benefits from user profiles in various areas. However, these profiles can also be used against users. Data brokers offer service providers access to customer behavior information, often without the customer's knowledge. Armed with this information, these service providers can then target customers with highly personalized ads, promotions, and even phone calls, all based on the data provided by the data broker. As digital technology and data collection become ubiquitous, such practices are becoming increasingly opaque. Users are often unaware of what personal data about them is collected, analyzed, and shared or sold. Small businesses may upload their customer data to data brokers, who combine it with online data collected in real-time from other sources.

A entire section of business has flourished that focuses on exploiting and monetizing user’s personal data. Users are often unaware of these companies as they do not interact with them directly. These data companies capture data from every intersection of the digital world, creating dynamic and fragmented profiles that are distributed across multiple databases.
Google engages in data surveillance by continuously tracking a user's various activities such as website visits, Gmail account usage, mobile device movements, and YouTube usage, among others, without the user being fully aware of its implications. This monitoring can result in the collection of highly personal data, as demonstrated by the presence of trackers on a bank website even after a user has logged in, as shown in Figure \ref{figure3}. Companies make no effort to improve the existing nontransparent play. The degree of corporate control over the current digital environment is a cause for concern with regard to information privacy. \cite{new1} investigated the transformative impact of digialization creating privacy challenges. Their analysis emphasized the need for robust solutions to safeguard patient information and pave the way for responsible, ethical adoption of digital electron healthcare. Profile data gathered by companies could be potentially abused for personal, organizational, or political purposes.

The ubiquitous sharing of personal data from apps, websites, devices, and services with third-parties and the subsequent use of this data by various companies has raised concerns. E-commerce companies rely heavily on distributed computing, which have their data centers spread globally. Data sets containing significant amounts of personal information about users are gathered, which can be used legitimately or exploited. For example, Amazon has operations across 12 regions, each with multiple data centers that are susceptible to physical attacks and persistent cyber-attacks. In 2014, eBay's customer data was compromised in a cyber-attack that resulted in the theft of names, email addresses, physical addresses, telephone numbers, birth dates, and encrypted passwords. The stakes are higher than ever due to the large number of people involved in personal data security incidents keeping privacy of user at stake. At Arkansas University, the professional development system was breached causing exposure of personally identifiable information of individuals \cite{r72}. Table \ref{tbl4} displays some high impact data breach incidents that occurred between 2021 and 2024. These incidents create opportunities for cyber-criminals to exploit vulnerabilities, such as through phishing, leading to severe consequences like unauthorized purchases, fund theft, or identity theft for individuals.
\begin{sidewaystable}[h]
\caption{\label{tbl4}{Some high impact data breaches reported between 2022 and 2024. [\cite{r153}, \cite{r130}, \cite{r154}, \cite{r155},\cite{new4},\cite{bb1}]}}
\begin{scriptsize}
\begin{tabular}{p{2cm} p{1cm} p{1cm} p{7cm} p{5cm}}
\toprule
\textbf{\textit{Victim}} & \textbf{\textit{Occurred}} & \textbf{\textit{Impact}}&\textbf{\textit{Cause/vulnerability}} & \textbf{\textit{Compromised data}} \\
\hline
\textbf{Texas Department of Insurance} & May 2022 & 1.8 million & \begin{itemize} \begin{scriptsize} \vspace{-2.5ex}
\item Security issue with a web application that manages workers’ compensation information. \item This vulnerability was due to programming code that allowed internet access to a protected area of the application. \end{scriptsize} \end{itemize}
& Social security numbers, addresses,dates of birth, phone numbers and information about worker's injuries.\\ 
\hline
\textbf{Twitter} & July 2022 & 5.4 million & \begin{itemize} \begin{scriptsize} \vspace{-2.5ex}
\item Vulnerability on the social media site. \item Allows to obtain Twitter ID without any authentication of any user by submitting a phone number/email \item Possible even though the user has prohibited this action in the privacy settings. \end{scriptsize} \end{itemize} &
Email addresses and phone numbers of celebrities, companies, randoms, etc.\\

\hline
\textbf{ChatGPT} & March 2023 & $\approx1\%$ data & \begin{itemize} \begin{scriptsize} \vspace{-2.5ex} \item Vulnerability in the code’s open-source library ``Redis". \item Used to cache user information for faster recall and access. \end{scriptsize} \end{itemize} & Credit card information and the titles of some chats they initiated. It was also possible to see another active user’s first and last name, email address, payment address, the last four digits (only) of a credit card number, and credit card expiration date. \\ 
\hline
\textbf{Norton Healthcare} & May 2023 & 2.5 million & \begin{itemize} \begin{scriptsize} \vspace{-2.5ex} 
\item  Unauthorized individual's access to the company’s network storage devices \end{scriptsize} \end{itemize} & Names, contact details, SSN numbers, dates of birth, health and insurance information, medical ID numbers, driver's license numbers, government ID numbers, financial account numbers, and digital signatures  \\
\hline
\textbf{Facebook Marketplace} & February 2024 &  0.20 million & Facebook Marketplace database was stolen using the `algoatson' Discord handle after hacking the systems of a Meta contractor. &  Names, phone numbers, email addresses, Facebook IDs, and Facebook profile information.\\
\bottomrule
\end{tabular}
\end{scriptsize}
\end{sidewaystable}

\subsection {Security Issues}
The insights obtained from the gathered data are of significant value to criminals. Illicit trading of individuals' personal information, such as their financial details, can assist fraudsters in locating their targets and lead to further harm through techniques such as fake calls and identity fraud. As per a study by \cite{r123}, nearly 47\% of Americans experienced financial identity theft in the year 2020, with the loss from identity fraud rising from $502.5$ billion in 2019 to $712.4$ billion in 2020, marking a 42\% increase. In 2014, an intruder took advantage of the vulnerabilities in InfoTrax's server and a customer's website by deploying a malicious code that enabled remote access to InfoTrax's server, allowing them to extract data from systems. By interrogating certain databases that contained full names, email addresses, phone numbers, addresses, Social Security Numbers (SSNs), admin IDs and passwords, and distributor user IDs and passwords, the intruder obtained the personal information of nearly one million customers. The attacker also obtained access to over 2300 unique full payment card information \cite{r98}. Such personal information, as in the case of the InfoTrax breach, can be exploited to carry out fraudulent activities. Stolen names, addresses, and SSNs could be used by identity thieves to apply for credit cards in the victim's name, which could affect the victim's credit score when the identity thief fails to pay the credit card bills. Understanding the financial and personal information of potential victims, including their income and other personal details, can aid criminals in committing severe crimes such as blackmailing or kidnapping. For instance, patients at Vastaamo clinic in Helsinki were blackmailed after their data was stolen in two breaches. Attackers demanded a €450,000 (\$530,000) bitcoin ransom and published clinical records on the dark web. Names, contact details, and therapy notes of hundreds were compromised \cite{r163}. User profiles make it convenient to find users that have similar behaviour, making certain section of the population vulnerable to criminal activities. For example, gathering profiles of young girls can enable criminals to morph their public photos and expose them to pornography. 

In 2016, an analytics company called ``Cambridge Analytica" gained unauthorized access to at least 87 million profiles of Facebook users, and utilized this data to create voter profiles to influence their voting decisions. This company provided voter data and analytics to both the Trump campaign and the campaign of former national security adviser John Bolton during the 2016 election cycle. This breach and exploitation of personal data was considered detrimental to democracy by various legislators \cite{r97}. Therefore, the abundance of user information increases the likelihood of attracting intruders to the system, which increases the threats.

The data exposed in public forum during a particular security incident can be combined with other data breaches to create detailed profiles of potential victims by particularly determined attackers. Using this information, they can conduct much more convincing phishing and social engineering attacks or even commit identity theft against those whose information has been exposed. In next section, we explain security policies in place and their challenges.

\subsection {Existing security policy and its challenges}
The right to privacy lacks a clear definition and legal protection, especially in the face of internet-driven data collection and sharing. New technologies pose fresh challenges to privacy, calling for a reevaluation of laws and safeguards in the digital age \cite{new2}. Many smart devices use a permission-based security model to prevent unauthorized access to sensitive system resources and user data. For instance, Android, which is among the most widely used operating systems for smart devices, requires application developers to request permission from users to access sensitive resources like GPS sensors, cameras, microphones, and private data via their Android manifest file \cite{r110}. This model gives users control over their device by allowing them to decide which apps can access their resources and data. However, this security mechanism is not entirely foolproof \cite{r111,r112}, as intruders can use two common techniques to bypass it: side channels and covert channel attacks, which exposes alternate mechanisms to access the system resources that unfortunately are not audited by the permission mechanism \cite{r108}. \textit{Side channel} attacks exploit security vulnerabilities inherent in the system to access system resources, for example, by leveraging the state of the CPU cache to infer sensitive data during program execution. The key insight here is that as programs execute, the state of the cache gets updated constantly. Considering the speed of CPU caches, these side channels can leak large amounts of data quickly. \textit{Covert channels} involve intentional efforts by two parties to share system resources without violating the underlying security mechanism, such as through shared storage. For example, a communication can involve directly or indirectly writing data to a storage location and another process directly or indirectly reading that location as in case of shared storage. Therefore, while security policies are in place, they may not provide sufficient protection for user data on their devices.

Maintaining user privacy is crucial when conducting profiling, therefore it is imperative to implement necessary security measures to safeguard personal information. There are policies being placed to protect user's data in many countries. The Digital Personal Data Protection Act 2023 of India, for example, aims to regulate the handling of digital personal data in a manner that respects both individuals' right to safeguard their personal information and the necessity to process such data for lawful purposes, as well as other related or incidental matters \cite{r165}. Next, we present an experimental analysis of data leakage from mobile phones using these identifiers and trackers.

\section{Case study of Privacy Leakage}
In a study by \cite{r126}, the exposure of Personally Identifiable Information (PII) from widely used applications on iOS, Android, and Windows platforms was uncovered. \cite{r127} identified PII exposure within HTTP packets through the classification of android apps packets. To classify outgoing HTTP packets, federated Learning was applied, in keeping with the data sensitivity and privacy concerns of the users. As of 2023, instances of PII exfiltration from mobile apps persist. This section presents a case study concentrating on identifying possible exfiltration of sensitive personal data from Android applications installed on smartphones.
\subsection{Clear text data transmissions in smartphones}

Mobile devices often need access to personal and crucial user information in order to function effectively and fulfill their intended purposes. However, this necessity can also expose potential privacy risks, as sensitive data becomes susceptible to unauthorized access, misuse, or exploitation. Mobile apps and the third-party libraries integrated within them might transmit PII to numerous external application servers, which require this data for service provision or user tracking. Encryption plays a vital role in ensuring privacy in digital communications. However, data packets transmitted without encryption can be easily intercepted by adversaries and network observers. Therefore, if data is transmitted from devices in plain-text format, it may be considered a significant design flaw. Even if plain-text data is compressed, it can still be effortlessly restored to its original form by recovering the compressed message and employing widely used compression algorithms. For instance, \cite{r115} identified clear text sensitive medical information, known as e-PHI, within encrypted communications. 

The National Institute of Standards and Technology provides a definition of PII as ``Any representation of information that permits the identity of an individual to whom the information applies to be reasonably inferred by either direct or indirect means'' \cite{r144}. The General Data Protection Regulation (GDPR) states that personal data should only be collected for explicit, legitimate, and specific purposes, and should not be processed in any way that is incompatible with those purposes. Numerous studies in the literature focus on controlling user tracking and preventing the unauthorized extraction of PII from mobile devices. These studies can be broadly categorized into three groups: permission-based analysis \cite{r146}, static and dynamic analysis of source code \cite{r147,r148}, and network-based analysis \cite{r126,r149,r138,r150}. In the following section, a technique is examined for gathering network traffic from smartphones and identifying any clear text data that might expose sensitive user information and behavior. The experiment entails two stages: collecting the traffic and detecting instances of plain-text sensitive information.

\subsection{Experimental Design}
\begin{figure*}[h]
\centering
\includegraphics[width=0.75\linewidth,height=0.23\textheight]{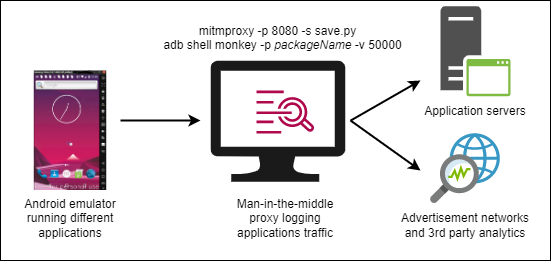}
\caption{\label{figure-7}Data capturing framework}
\end{figure*}

Figure \ref{figure-7} illustrates the design of our network data capturing framework for Android devices. The network data capturing and monitoring system comprises a smartphone and a man-in-the-middle framework or tool. We perform the experiments on an emulator, and the analysis of iOS applications remains a part of our future work. We utilize Genymotion emulator to simulate mobile devices due to their wider variety and faster, lightweight nature compared to other frameworks. To intercept the traffic of mobile devices, we use Mitmproxy, a command-line tool that acts as an HTTP/HTTPS proxy and records the traffic.

Mitmproxy connects the smartphone to the Internet, therefore recording all incoming and outgoing traffic. As long as the client trusts mitmproxy's built-in certificate authority, Mitmproxy will decrypt the traffic. Most of the time, this means installing the mitmproxy CA certificate on the client device. So, the certificate is added to the trusted store of the emulator. For this the device (emulator) is rooted. In this experiment, the applications mentioned in Table \ref{tbl5} that have at least 10 million downloads are selected. By doing so, it is ensured that we are analyzing the most popular applications used by the majority of users as of March, 2023. These applications are downloaded from the playstore and we manually sign up and log in to the application because the tools that generate automatic UI events lack this functionality. UI/Application Exerciser Monkey is used to generate synthetic user inputs and does a 10 minutes tour of the application. The traffic captured undergoes preprocessing by employing a set of decoding operations utilizing the UTF-8 scheme in order to store the traffic in the appropriate format in log files for final analysis.

Before running the application, we grant it all the permissions it requires. We then execute each application while capturing the network traffic it generates. In our tests, certain applications generated no traffic either because they detected our device was rooted or because they required a SIM card or valid phone number that our test device lacked. Therefore, we removed these types of applications from our list and analyzed the remaining ones.
\begin{figure*}[h]
\centering
\includegraphics[width=0.90\linewidth,height=0.25\textheight]{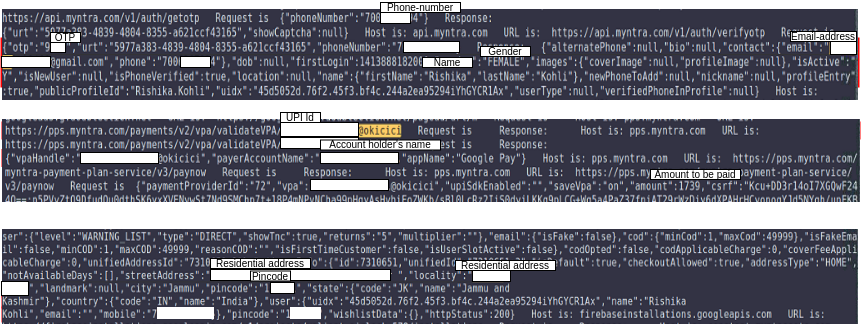}
\caption{\label{figure8}Leaks captured from Myntra application (\textit{Version: 4.2201.1; Accessed Date: March, 2023})}
\end{figure*}
\subsection{Collected Dataset and Results}
In order to gain a comprehensive understanding of mobile tracking and advertising, we chose to diversify our app categories. Therefore, we run a total of 18 applications under 6 categories: \textit{Social, Education, Entertainment, Travel, Shopping and Others}. Let \textit{D\textsubscript{t,A}} be the data transmitted at time \textit{t} from Mobile application \textit{A}. We manually check \textit{D\textsubscript{t,A}} for the presence of three parameters: location, user-identifier, and device identifier. For instance, Figure \ref{figure8} depicts the user's details such as his name, gender, phone-number, OTP required for login,  email-address, UPI ID and residential address are exposed in plain-text from Myntra application. The discovered leaks of privacy have been summarized in Table \ref{tbl5}. Among various app categories, social and shopping apps exhibit the highest degree of user location leakage, while user gender and date of birth are commonly exfiltrated user's identifiers across all app categories. Furthermore, the exposure of device identifiers reveals the pervasive tracking ecosystem prevalent within the applications. Results show that a substantial amount of user information can be extracted from network traffic and is mostly hidden from users. We have opted for crowdsourcing to label the dataset for the presence of PII. Recognizing human judgment as crucial in dataset evaluation, we engaged 5 users in labeling the dataset, and the task is nearly 60\% complete.
\subsection{Privacy Inference Analysis}
The purposes of collecting data from various applications can be classified as either obvious or non-obvious. The obvious purpose of data collection is to ensure proper app functionality for users, which can be further divided into the categories of ``benign'' and ``non-Benign''. For instance, in the case of Skype, it is obvious and considered benign to transmit a user's name from the device for the application to function, while transmitting and revealing a user's text messages is non-benign and violates privacy. Non-obvious purposes of data collection can also be categorized as ``Benign'' and ``Non-Benign''. Apart from collecting data for the primary app function, some apps collect more data and permissions than necessary for other purposes. For example, Skype's transmission of a user's static location provided during profile creation is considered benign, whereas transmitting a user's phone number is deemed non-benign.

Many apps offer convenient login options that allow users to sign in using their Google or Facebook credentials. Although this simplifies the login process, it also means that the app receives the user's Google or Facebook account information. LinkedIn traffic analysis has revealed that certain details, such as a user's Google profile picture, are shared and transmitted through URLs. While these leaks may not initially appear to be dangerous, they provide enough data for potential eavesdroppers to create a profile of the user.

Adversary can build a profile of a user using temporal expansion based on the aggregated data. Although it appears that at t=t\textsubscript{0}, adversary can infer very little information about user. But over time $t=t\textsubscript{0}+\Delta t\textsubscript{1}$, it is possible to infer more details about the user. Finally $t=t\textsubscript{0}+\Delta t\textsubscript{1}+\Delta t\textsubscript{2}$ gives a detailed look into the life of a user. Spatial expansion is another aspect of generating a user's profile, involving the concatenation of data from various applications into a unified profile. Let's consider the data gathered by an adversary from a Skype application, represented as d\textsubscript{skype}. By analyzing data from applications like Cleartrip (d\textsubscript{cleartrip}) and Myntra (d\textsubscript{myntra}), the adversary can create a more comprehensive profile d=d\textsubscript{skype}+d\textsubscript{cleartrip}+d\textsubscript{myntra} of the user's behavior. For instance, in a threat model, the adversary may use d\textsubscript{skype} to track the user's current location, d\textsubscript{cleartrip} to determine their vacation plans, and d\textsubscript{myntra} to find the user's residential address. This type of information could potentially be exploited by an observer for criminal activities such as burglary when the user is away on vacation. Emphasizing the significance of strengthening user data privacy becomes apparent in this threat model, aiming to prevent similar situations from occurring in the future. For instance, \cite{r116} studied the feasibility of fingerprinting motion sensors. Authors identified the type of smartphone based on the data transmitted by accelerometer and gyroscope. It was concluded that it is feasible to fingerprint smartphones from its motion sensors and can be accessed by web page publisher or advertisers without users’ awareness. 

\begin{table*}[!h]
\caption{\label{tbl5}Summary of private data leaks in Mobile App Traffic (\textit{Duration: Oct 2022 - March 2023}). $^3$These are versions of the applications that have been analyzed during our experiments. $^4$ This is as per Indian playstore and here M represents Million and B represents Billion.}
\begin{scriptsize}
\resizebox{\textwidth}{!}{%
\begin{tabular}{p{1.5cm}|p{1.5cm} p{1.5cm} p{1.5cm} p{1.5cm} p{4.0cm} p{5.2cm}} \\
\toprule
\textit{\textbf{Category}} & \textit{\textbf{App}} & \textit{\textbf{App version\footnotemark[3]}}& \textit{\textbf{No. of downloads\footnotemark[4]}} & \textit{\textbf{Location}}& \textit{\textbf{User Identifiers}}& \textit{\textbf{Device Identifiers}}\\
\hline
\multirow{3}{*}[-10pt]{Social} & Pinterest & \textit{7.43.1} & 500M+ & \centering{-} & First name, last name & IP address, device configurations and details, Google advertising ID, application install and last update time\\
\hhline{|~|------|} 
 & Skype &  \textit{8.81.0.268} & 1B+ & Location of user and also his contacts & Name, gender, date-of-birth, email-id, text messages, phone number, ID of user and his contacts  & Device model name\\
\hhline{|~|------|} 
 & LinkedIn & \textit{4.1.710} & 1B+ & \centering{-} & First name, last name, profile picture URL, headline of profile, session-key and session-password & Device identifiers (Android ID)\\
\hhline{|~|------|} 
& WhatsApp & \textit{2.22.13.76} &5B+ & \centering{Country} & Phone number, email-id
& Device details (device name, OS, android version), Android ID, device ID, SDK version\\ 
\hhline{|~|------|} 
&  Reddit &\textit{2020.8.2} &100M+ & \centering{Country} & User ID, user name, user password, email-id
& Device details (Android version, device name, OS, OS version, hardware ID, device fingerprint id, device dimension), advertising ID, application install and last update time, device ID, SDK version.\\
\hline
\multirow{1}{*}[-6pt]{Education}
 & Gradeup & \textit{11.07} &10M+ & \centering{-} & Name, phone number, OTP used while registering, email-address  & Device identifiers and configuration\\
\hline
\multirow{2}{*}[-6pt]{\begin{tabular}[c]{@{}c@{}}Entertain-\\ment\end{tabular}} & Voot & \textit{4.2.7}& 100M+ & \centering{-} & Gender, age, phone number, email-id & Device details, Google advertisement ID\\
\hhline{|~|------|} 
& Zee5 & \textit{35.1338119.0} & 100M+ & \centering{Country, region} & Email-ID, first name, last login, gender, date of birth, phone number & Device details (Android version,device name), device ID, install time.\\
\hline
\multirow{1}{*}[-20pt]{Travel} & Cleartrip & \textit{22.3.0} & 10M+ & \centering{-} & Username, email-address, password, phone number, date-of-birth, gender, registration details (check-in date, check-out date, number of adults and children, destination city and country)
& Device identifiers, device details (model, network, device dimensions, network type, CPU type, OS version, screen dpi), advertising IDs.\\
\hhline{|~|------|} 
& OYO &\textit{5.9.2}&50M+ &  \centering{-} &Name of user, date of birth, gender, marital status, phone number, email-id & Device identifiers and google advertising ID. \\
\hline
\multirow{2}{*}[-20pt]{Shopping} & Dominos India & \textit{9.8.18} & 50M+ & User’s address & First name, last name, mobile number, email-id & Device identifiers, device details (model, network, device dimensions, network type, CPU type, OS version, screen DPI), advertising IDs\\
\hhline{|~|------|} 
 & Myntra & \textit{4.2201.1} & 100M+ & User’s address & First name, last name, phone number, gender, first login date, OTP for login, UPI ID, email-id & Device details, IP address, network type\\
\hhline{|~|------|} 
&  eBay & \textit{6.49.0.3} & 100M+ & \centering{Country} & Email ID & Device details (model, dimensions, OS, hardware, manufacturer, OS version, physical memory,processor, architecture, RAM, disk space, processor count, processor word size, system up time, thermal state, time zone, user language)\\
\hline
\multirow{2}{*}[-16pt]{Others} & Airtel Thanks & \textit{4.40.10} & 100M+ & Location (not exact) &  Name, phone number, alternate number, email-id & siNumber, device details, advertisement IDs, local IP.\\
\hhline{|~|------|} 
& Adobe Acrobat Reader & \textit{22.6.0.22829} & 500M+ & Latitude and longitude & First name, last name, email-address, contents of the PDF opened in application, list of files in Google drive & Device ID, OS, device details, device configurations, connection type, local IP, advertising ID. \\
\hhline{|~|------|} 
 & Flipboard & \textit{4.2.93} & 500M+ & \centering{-} & Name, email-id  & Device brand\\
\hhline{|~|------|} 
& Truecaller & \textit{12.17.8} & 500M+ &\centering{-} & Phone number and email-id
& Device ID, device details (OS, Model) and sim number\\
\hhline{|~|------|} 
&  Zoom & \textit{}& 500M+ & \centering{Country} & Phone number & Device details (model, dimensions, OS), Device ID, SDK version. \\
\bottomrule
\end{tabular}}

\end{scriptsize}
\end{table*}
\subsection{PII Detection in network traffic}
There are many approaches that can be used to analyze the network traffic of mobile devices in an efficient and secure manner in the existing state-of-the-art. \cite{r126} used machine learning in order to train C4.5 Decision Tree for detecting leaks in a centralized manner. Our work \cite{my} analyzed the dataset provided by ReCon and provide improved versions of classifiers inorder to enhance detection framework.

\subsubsection{Dataset Overview and Pre-processing}
In ReCon's work, controlled experiments are conducted using smartphones, followed by factory resets and connecting the devices to Meddle \cite{r139}. With Meddle, all network traffic is routed through a proxy server using VPN tunnels. Traffic is intercepted and modified by software middleboxes once it reaches the proxy server. Their raw IP traffic is logged using tcpdump and HTTP flows are extracted using bro. In the following steps, they look for PII loaded onto devices that are conspicuous.

Algorithm 1 shows the feature extraction and selection technique: We utilized a bag-of-words model inspired by ReCon's approach.   This model treats certain characters (such as , ; [] / ()) as separators, defining words (feature) as sequences of characters between them. A binary vector is created, marking a word with an integer greater than 1 based on its occurrence in a flow, or with 0 otherwise. However, the model generates an extensive output, and several steps are taken to reduce and select relevant features. Initially, a feature is eliminated if its word frequency is below a predefined threshold. To preserve rarely occurring PII, oversampling is conducted to surpass the selected threshold. To prevent the model from utilizing PII values as features, randomization of PII values in each flow during training is done. Lastly, stop-word-based filtering using tf-idf is employed to eliminate commonly occurring words in HTTP flows (e.g., content-length, en-us, etc.). Only features with low tf-idf values, which did not appear adjacent to a PII leak in a flow, are considered.

\subsubsection{Methodology}
\begin{figure}[!h]
\centering
\includegraphics[width=\linewidth,height=0.28\textheight]{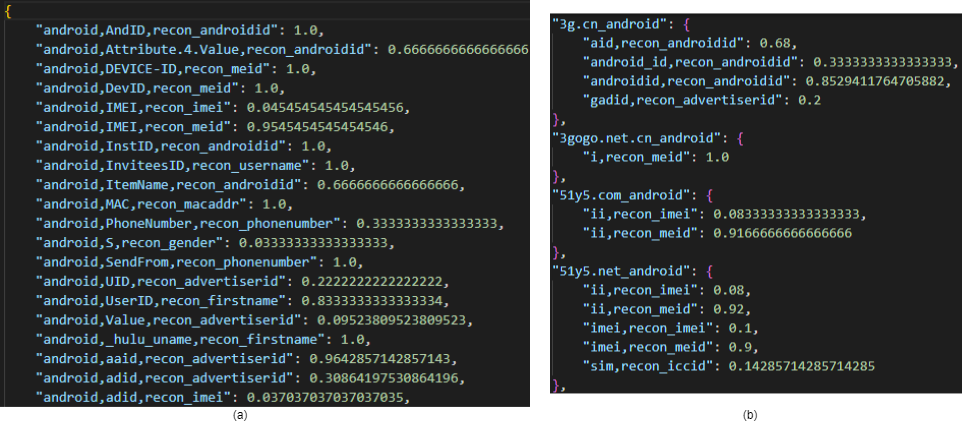}
\caption{\label{figure6} a) Shows list of probabilities that particular keyword corresponds to a PII value. b) Shoes list of probabilities calculated for keys in each domain and OS.}
\end{figure}

To start our work we first pre-processed the dataset to select the relevant features using tf-idf. As mentioned earlier, following the elimination of frequently appearing words in HTTP flows using tf-idf, we proceed to iterate through all features. Probabilities calculated to map a particular keyword to PII value is depicted in Figure \ref{figure6}. For each feature, we calculate the confidence level using heuristics, similar to the approach in the ReCon's work.

\vspace{-0.2cm}
\begin{equation*}
  Heuristic 1:  P_{type,key}=\frac{K_{PII}}{K_{all}}
\end{equation*}

\begin{figure*}
\centering
\includegraphics[width=0.85\linewidth,height=0.30\textheight]{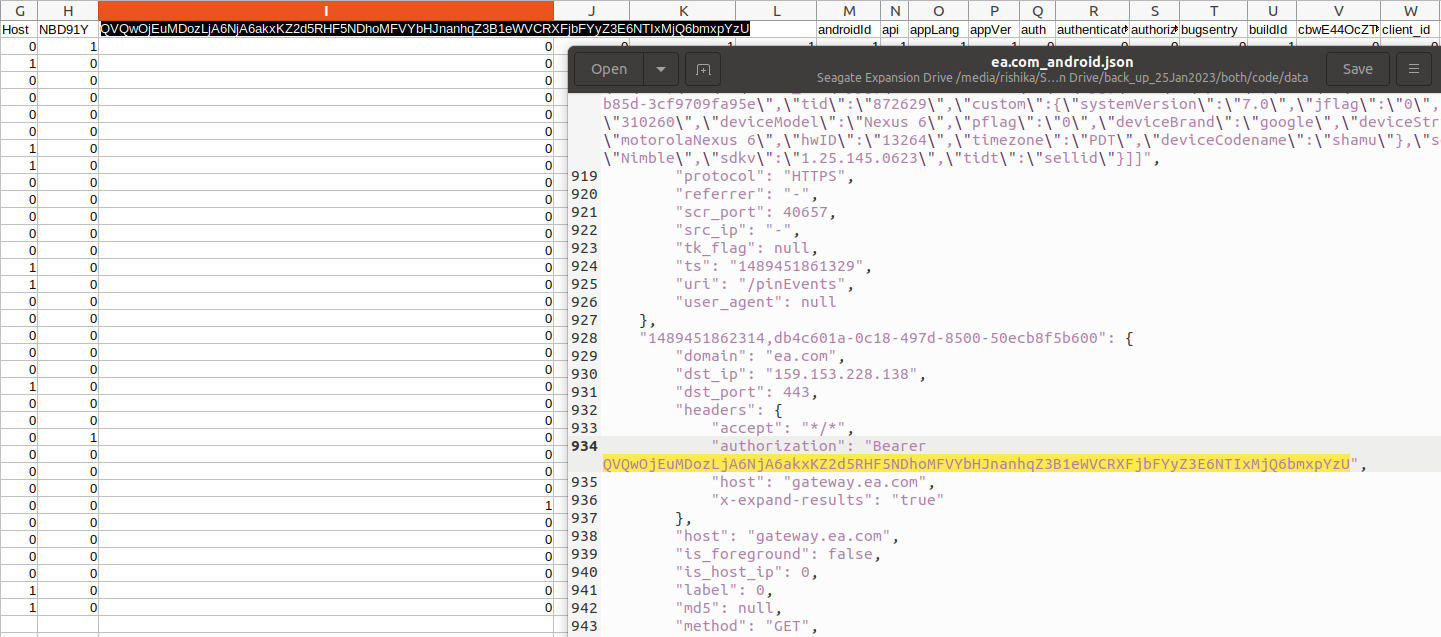}
\caption{\label{figure7} Features extracted using Algorithm 1 for a single domain.}
\end{figure*}

Here, $K_{PII}$ gives number of times the key appeared in flows with positive PII leakage and $K_{all}$ gives number of times the key appeared in all flows. Features having confidence level greater than chosen threshold are selected for further examination. Figure \ref{figure7} shows a snap-shot of features selected for a domain named “ea.com” for android OS.
\begin{algorithm}[!h]
\begin{scriptsize}
\DontPrintSemicolon
\SetAlgoLined
\SetKwInOut{Input}{Input}
\SetKwInOut{Output}{Output}
\Input{a) Threshold for word frequency: $freq_t$,\\
b) Stop-word list: $stop\_words$,\\
c) Threshold for tf-idf value: $tfidf_t$}
\Output{Selected features for training the model}
\BlankLine
\begin{algorithmic}
  \STATE Define list of separators:\\
$separators \gets \{\text{`,', `;', `\{\}', `[]', `/',\dots, `()'}\}$\\
Initialize a 2-D matrix $V$.

  \FOR{each HTTP flow $i=1,2,\dots,n$}
    \STATE $W_i \gets \text{Split}(flow_i,separators)$
  \ENDFOR
  
  \FOR{each word $w_j \in W_i$ where $1 \le i \le n$}
      \IF{$w_j$ appears in $flow_i$}
        \STATE mark $V_{i,j}$ as 1
      \ELSE
        \STATE mark $V_{i,j}$ as 0
      \ENDIF
  \ENDFOR
  
  \STATE Randomize any PII values in $W_i$ to obtain a new set $W_i'$
  
  \STATE Calculate the word frequency for each word $w_j$ across all flows: $f_j=\sum\limits_{i=1}^n V_{i,j}$

  \FOR{each word $w_j$}
    \IF{$f_j < freq_t$ \textbf{and} $w_j \notin \text{PII}$}
      \STATE Remove $w_j$ from the list of features.
    \ENDIF
    \IF{$f_j < freq_t$ \textbf{and} $w_j \in \text{PII}$}
      \STATE Oversample $w_j$ to meet the threshold.
    \ENDIF
    \STATE Calculate the tf-idf value for each word $w_j$ across all flows.\\
  \ENDFOR

  \IF{$tfidf_{i,j} > tfidf_t$ \textbf{or} $w_j \in stop\_words$}
    \STATE remove $w_j$ from the list of features.
  \ENDIF

  \FOR{each $flow_i$}
    \STATE Remove all features that appear adjacent to a PII leak in $W'_i$.
  \ENDFOR

  \STATE Output the selected features as a set of binary vectors $V_{i,j}$ for each $flow_i$.
\end{algorithmic}
\caption{Feature extraction and selection}
\end{scriptsize}
\end{algorithm}

To initiate our experiments, we partitioned the ReCon dataset into two sets: a training set containing 60 domains and a testing set with 12 domains. The training set comprises approximately 7200 data inputs and 6500 features. Our initial step involved utilizing a Decision Tree to identify sensitive information in network traffic packets. Our goal is to evaluate and improve the performance of the Decision Tree model compared to previous works in detecting PII. We experimented with feature manipulation, removing certain features from the dataset using various heuristics. Detecting PII in network flows often requires recognizing intricate patterns and relationships among different data points. Leveraging the ability of neural networks to model non-linear relationships, we decided to apply a neural network model to this task after the Decision Tree experiments. Our neural network model employs the ReLU activation function in all hidden layers to address vanishing gradient issues during back-propagation. For our binary classification problem, the output layer uses the sigmoid activation function to ensure the output $Y_i$ falls within the range $[0,1]$. The loss function $L$ employed in the current context is the binary cross-entropy function.
\vspace{-0.2cm}
\begin{equation*}
    L = -\frac{1}{n}\sum_{i=1}^{n}(Y_i \cdot log\hat{Y_i} + (1-Y_i) \cdot log(1-\hat{Y_i}))
\end{equation*}
 RMSprop optimizer is used to minimize $L$. At iteration $t$ for weight $w$, RMSprop update is given by:
\vspace{-0.2cm}
\begin{equation*}
v_t=\gamma*v_{t-1}+(1-\gamma)*\left(\frac{\partial L}{\partial w}\right)_{t}^{2}
\end{equation*}
where,
$v_t$ is the moving average of squared gradients at $t^{th}$ iteration, $\gamma$ is decay rate controlling the weights given to previous squared gradients in the moving average. Therefore, weight update follows:
\vspace{-0.2cm}
\begin{equation*}
w_{t+1} = w_t - \frac{\alpha\left(\partial L/\partial w\right)}{\sqrt{v_t}+\epsilon}
\end{equation*}
where,
$\alpha$ is the learning rate and $\epsilon$ is a small constant typically set
in the range of $10^{-8}$ to $10^{-10}$. The choice of $\alpha$ determines the time needed for model to converge. If it is fixed too small, it slows the convergence, on the contrary, a large value of $\alpha$ will possibly result in oscillation, preventing the error to decrease below a certain value.

In our study \cite{spcom}, upon examination, we found numerous features that are repetitive or duplicate. For instance, features like `connection' and `Connection' are regarded as distinct features despite having same meaning. Such duplication could adversely impact the effectiveness of the detection framework.

Moreover, we observed that certain encoded strings, such as

\begin{tabular}{p{\linewidth}}
``QVQwOjEuMDozLjA6NjA6akxKZ2d5RHF5NDhoMFVYbHJnanhqZ3B1eWV
CRXFjbFYyZ3E6NTIxMjQ6bmxpYzU''
\end{tabular}

\noindent are incorporated as features, yet they lack meaningful information. Including these strings as features merely inflates the total number of features without adding any value in terms of informative content. Hence, we implemented various techniques to filter out such features, as elaborated below:
\begin{enumerate}
    \item Removing duplicates: Removing duplicates involved a two-step process. Initially, we transformed all feature names to lowercase, ensuring uniformity. Subsequently, we grouped similar features together, effectively eradicating redundant columns from the dataset.
    \item To decode encoded strings and extract insights, we employed a tool named CyberChef \cite{cyber}, which employs diverse decoding operations to unveil the underlying information. Despite employing several decoding operations, it became apparent that many encoded strings either failed to reveal meaningful information or did not pertain to any PII. Consequently, we excluded such features from further analysis.
    \item Stemming: We employed Porter's Stemmer algorithm for suffix stripping to reduce features to their root forms, a process known as stemming. Following this, we clustered together all features that shared a common root form. This approach enabled us to streamline and organize the features according to their linguistic similarities. For instance, features `attempting' are grouped under feature `attempt'.
    \item We noticed that some features had lengths significantly longer than typical words, while others were very short. We theorized that PII words would likely fall within a certain length range and conform to English language patterns. For example, URLs or file names often have longer lengths. After experimentation, we found that setting a threshold of 70 effectively captured all relevant words. Hence, any features exceeding this length were omitted from further analysis. Additionally, strings with a length of 1 were also excluded.
\end{enumerate}

After filtering out the relevant features, all experiments were rerun on the refined dataset. The results of each experiment are detailed in the subsequent section.

\begin{figure*}[!h]
\centering
\includegraphics[width=\linewidth,height=0.23\textheight]{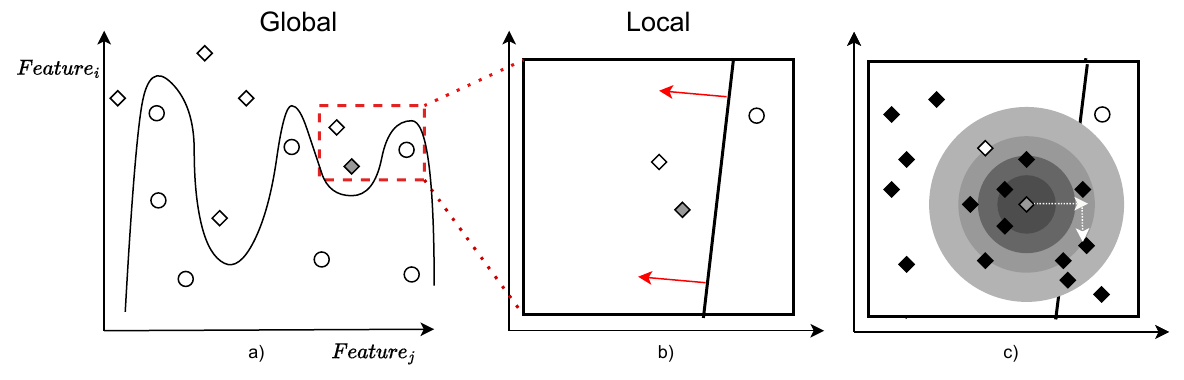}
\caption{\label{limeexp}LIME explaining the prediction of model $f(\cdot)$ a) Shows the global context and decision boundary created by complex non-linear model $f(\cdot)$. Grey point is the predicted instance for which explanations are needed. b) Shows the decision boundary for simple linear model $m(\cdot)$. Idea here is to see neighbourhood and do simple explanation of local region. Features that are important in global context may not be important in local area. c) Shows the perturbed data-points (black points). The exponential kernel ($\pi_x$) can be considered as a heat map. The perturbed data-points are weighted according to distance to the predicted instance}
\end{figure*}

To evaluate the interpretability of our best-performing models (decision tree and neural network), we utilized LIME  (Local Interpretable Model-Agnostic Explanations) \cite{LIME}. The main idea behind LIME is that a model is trained to approximate the predictions of the underlying black-box/complex model. LIME generates a new dataset consisting of perturbed samples and the corresponding predictions of the complex model as shown in Figure \ref{limeexp}. On this new dataset, LIME then trains an interpretable model $m$ such as linear model or decision tree that is weighted by the proximity of the sampled instances to the instance of interest. Let $\vec{x} \in \mathbb{R}^d$, represent an instance being explained. The weighted loss is defined as:
\begin{equation*}
\mathcal{L}(f,m,\pi_x) = \sum_{\vec{p} \in P} \pi_x(\vec{p}) \left( f(\vec{p}) - m(\vec{p}) \right)^2
\end{equation*}

The function $f$ represents the complex model being explained, with $f : \mathbb{R}^d \rightarrow \mathbb{R}$. $P$ represents the set of all perturbed instances, each having a weight $\pi_x$. $\pi_x(\vec{p})$ is the proximity measure between an instance $\vec{p}$ and $\vec{x}$, to define local neighborhood of $\vec{x}$. It is therefore, sum of squared distance between predictions of complex model $f(\vec{p})$ and predictions of simple model $m(\vec{p})$. 

\subsubsection{Results}
\begin{table}[!h]
\begin{minipage}[t]{0.40\linewidth}
\centering
\caption{Best results obtained.}
\begin{scriptsize}
\begin{tabular}{l|cc} 
\hline
\textbf{Model} & \textbf{Train} & \textbf{Test} \\
\hline
$DT_{\text{post\_heuristics},\uparrow \text{testing samples}}$ & 93.79\% & 72.89\% $\star$ \\ 
\hline
$NN_{\text{post\_heuristics},\uparrow \text{testing samples}}$ & 95\% & 74.5\% $\star\star$ \\ 
\hline
\end{tabular}
\end{scriptsize} 
\label{Table6}
\end{minipage}
\hfill
\begin{minipage}[t]{0.8\linewidth}
\centering
\caption{Summary of results obtained using LIME}
\begin{scriptsize}
\begin{tabular}{lccc} 
\hline
\textbf{Model} & \textbf{Train} & \textbf{Test} & \textbf{Tr. time (s)} \\
\hline
$DT_{\text{best},\text{w/o\_LIME}}$ & \textbf{93.79\%} & \textbf{72.89\%} $\star$ \footnote{\tiny Taken from Table 7 (row having $\star$ marker)} & \textbf{0.095} \\ 
\hline
$DT_{\text{with\_SHAP}}$\cite{my} & 99\% & 67\% & 0.16 \\ 
\hline
$DT_{\text{with\_LIME-H1}}$ & 92.6\% & 71.2\% & 0.004 \\ 
\hline
$DT_{\text{with\_LIME-H2}}$ & 91.86\% & 71.2\% & 0.005 \\ 
\hline
$NN_{\text{best},\text{w/o\_LIME}}$ & 95\% & 74.5\% $\star\star$ \footnote{\tiny Taken from Table 7 (row having $\star \star$ marker)} & 16.54 \\ 
\hline
$NN_{\text{with\_SHAP}}$\cite{my} & 84.7\% & 75.32\% & 5.45 \\ 
\hline
$NN_{\text{with\_LIME-H1}}$ & 91.01\% & 73.38\% & 3.13 \\ 
\hline
$NN_{\text{with\_LIME-H2}}$ & \textbf{95.29\%} & \textbf{75.01\%} & \textbf{3.615} \\ 
\hline
\end{tabular}

\end{scriptsize}
\label{Table8}
\end{minipage}
\end{table}
\begin{enumerate}
\item \textbf{Decision tree}:  We employed a similar decision tree (DT) model as proposed by ReCon on their dataset. Applying the same model to our filtered dataset resulted in a 62\% accuracy on the test data, indicating that feature filtering did not yield improvements compared to our previous work \cite{my}. After removing domain and OS features, the classifier's performance improved, but feature filtering alone did not prove beneficial. Incorporating \textit{heuristics-1} and setting a threshold to 0.2, as in \cite{my}, reduced the features from 6,538 to 3,343 but led to a decline in accuracy. The same model on the filtered dataset yielded a similar accuracy.

We then expanded our test samples, as in \cite{my}, while employing \textit{heuristics-1}. This approach improved the model's performance on the filtered dataset to 68\%. This increase may be attributed to the test sample points now demonstrating a distribution similar to that of the model's training data points. However, the issue of overfitting persisted. To address this, we applied cost complexity pruning to the model, which raised the accuracy to 72.89\% as shown in Figure \ref{plt1}. It shows how accuracy of the decision tree classifier changes with different levels of pruning (controlled by cost complexity parameter, `ccp\_alpha') for both the training and test datasets.

\begin{figure}[!h]
\centering
\includegraphics[width=0.55\linewidth,height=0.23\textheight]{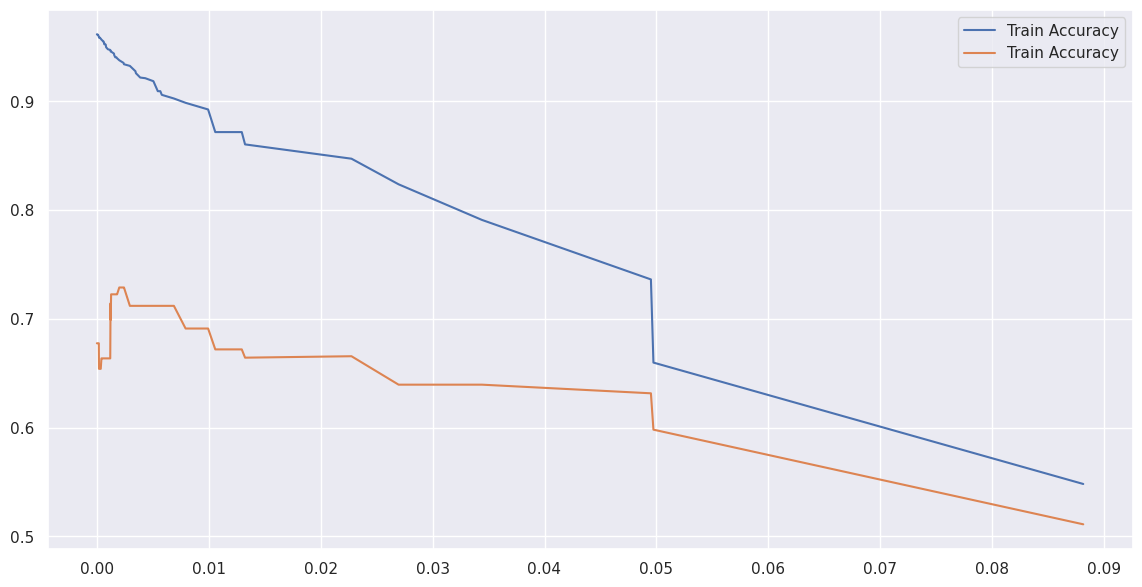}
\caption{\label{plt1}Performance of decision tree after using cost complexity pruning. Here, x-axis represents the values of the cost complexity parameter and y-axis represents the corresponding accuracy scores}
\end{figure}

\item \textbf{Neural network}:We constructed a neural network (NN) model with 1 input layer and 6 hidden layers, each containing 2048, 1024, 512, 256, 128, and 64 neurons respectively. The output layer comprised 1 neuron for binary classification. Following the approach of \cite{my}, we integrated 9 top features (including domain and OS) selected via Chi-Square scoring into the model. This yielded a training accuracy of 52.6\% and a testing accuracy of 62\%, mirroring the results of \cite{my}. Analysis of the NN's weights before and after training indicated the model's inability to effectively learn. 

Subsequently, using 7 features (after excluding domain and OS from the initially selected features) instead of 9, the NN maintained the same accuracy of 62\% as in \cite{my}. However, unlike the DT model, which achieved 67\% accuracy, this discrepancy suggested that the NN model wasn't learning optimally. 

Following the implementation of \textit{Heuristic-1}, the training accuracy surged to 92\%, but testing accuracy after feature filtering showed improvement to 57\% compared to \cite{my}. Nonetheless, these results suggested overfitting on the given dataset. To address this issue, we reduced the model's complexity from 6 to 3 hidden layers, akin to \cite{my}. While the training dataset exhibited a 95\% accuracy, testing revealed a 74.5\% accuracy, which is still 2.5

Table \ref{Table6} shows results of the best performing classifiers and indicate a significant enhancement in the performance of models as a result of filtering the features.

\item \textbf{Explainablity of obtained results after filetring using LIME}: 
We utilized LIME to refine the top classifiers in both categories by identifying crucial features using the following heuristics.

Given a dataset of $n_{val}$ validation samples (subset that helps to tune hyperparameters and assess the model's performance on data it hasn't seen during training) and $d$ features, let $imp_{i,j}$ represent the importance value of feature $j$ in sample $i$, where $i=1$ to $n_{val}$ and $j=1$ to $d$.  Initially, for each sample $i$, the set $imp_{i,nonzero}=\{imp_{i,j} | imp_{i,j} \neq 0$ for $j=1$ to $d\}$ is defined as the subset of non-zero feature importance values. These non-zero importance values are then sorted in descending order to form $imp_{i,sorted}$. 
\begin{itemize}
\item \textit{Heuristic 2} (H2): A $threshold$ is calculated as $Percentile(|imp_{i,sorted}|_{i=1}^{n_{val}},75)$. For each sample $i$, feature $j$ is retained if $|imp_{i,j}|$$>$$threshold$ and included in $selected\_features_i$. Finally, $final\_selected\_features= \bigcap_{i=1}^{n_{val}}selected\_features_i$
\item \textit{Heuristic 3} (H3): Let $\mathbf{a}_i=\left\lceil 0.2 \times d \right\rceil$ be the number of features chosen for sample $i$, where 0.2 means selecting the top 20\% of features, and $\text{top\_feature}_i$ be the set of top $\mathbf{a}_i$ features selected for sample $i$. The set $Common\_features=\{\text{top\_feature}_i \mid \text{count}_{\mathbf{a}_i} \geq 0.75 \times n_{val}\}$ contains the features common to 75\% or more of the validation samples.
 
\end{itemize}
Once the essential features contributing to classifier predictions were identified, the classifier was trained using only these selected features.

Using the LIME Tabular Explainer \cite{limetab}, we generated explanations for 200 validation samples, covering all 2202 features. Employing \textit{Heuristics-2 }and \textit{3}, we selected features of higher importance and retrained the best classifiers from Table \ref{Table6}. The results for each scenario are presented in Table \ref{Table8}. Remarkably, the NN classifier, utilizing \textit{Heuristic 2}, demonstrates comparable performance to the classifier in \cite{my} using SHAP, considering both training and test accuracy along with training time. While there is minimal impact on classifier accuracy with reduced features (38 in \textit{H2} and 124 in \textit{H3}), training time is reduced compared to the variations proposed in \cite{my} using SHAP. Conversely, the DT classifier doesn't benefit from LIME, as decision trees are inherently interpretable, and LIME may not provide significant additional value in this context.
\end{enumerate}
\subsubsection{Discussions}
The results indicate that eliminating unnecessary features improves the performance of machine learning models in predicting outcomes. Initially, our experiments followed ReCon's approach, where the DT model achieved a 62\% accuracy on test data both before and after filtering. Subsequently, we tested an NN model, initially matching the DT's accuracy. However, after removing two features from the dataset before training, the DT's accuracy improved from 62\% to 67\%, and post \textit{Heuristic-1}, the DT model exhibited signs of overfitting. In contrast, the NN model maintained a 62\% accuracy, failing to adapt or learn due to unchanged model weights. We then employed \textit{Heuristics-1} to enhance our training accuracy, but this resulted in overfitting. To address this, we simplified the model's complexity to mitigate overfitting. Consequently, our testing accuracy for the NN model increased from 47\% to 72\% after this adjustment, as in \cite{my}, and from 57\% to 74.5\% after post-filtering. On the other hand, applying pruning to the DT model yielded an accuracy of 72.89\% on testing data. After applying LIME for feature selection, the accuracy increased to 75.01\%, with a reduced training time of approximately 3.62 seconds for the NN. However, the DT model without LIME provided superior results.

\section{Open research directions}
Extensive research has been conducted on user profiling; however, a disparity remains between the present technology and future requirements. The most demanding elements of the user profiling process involve establishing profiles for new users and consistently updating existing users' profiles to align with their evolving needs, interests, and preferences \cite{r4}. Therefore, the primary hurdle in developing an adaptive personalized application lies in constructing an automated user profile.


We highlight several open research areas aimed at enhancing user privacy through the implementation of security measures.

\begin{enumerate}
\item \textit{Could user information be shared with third parties in an unclonable format? Can an unclonable transformation be applied to user information before sharing it? Is it possible to share user information with first-party domains and then with third-party domains in a transparent manner, without hiding it from users?}

If a user declines to share information with external domains, it should be prohibited from leaving an internal system and being transmitted to the outside world. However, if the user needs to share certain information with a domain in order to access its services, it can be shared in a format that cannot be replicated or cloned. Before transmitting the data from the user's device, a transformation can be applied to prevent the creation of replicas of the user's data.

\item \textit{How can users be informed and provide consent to first-party entities who have collected their data to share it with third parties?}

Is it possible to establish a system that can track the movement of user data regardless of where it travels. Once a user has given consent to share personal information with any domain, those domains should refrain from selling their customers' data to external third parties. If they do, users should receive notifications.



\item \textit{What are the possibilities of sharing information via smart contracts?}

The increasing popularity of smart contract technology is due to its advantages, including security, transparency, cost-effectiveness, and autonomy. Essentially, smart contracts consist of predefined rules agreed upon by involved parties. Such a system could be advantageous in the context of data sharing. When users and the domains they interact with agree on conditions related to data sharing, users gain greater control over their personal information.
\item \textit{Can role-based access be provided to different requesting domains by end users?}

All first-party and third-party domains, as well as trackers present in applications that users interact with, can access data transmitted by users' devices. In this scenario, implementing role-based access systems can be beneficial by granting access to requesting domains based on specific conditions. For instance, a user can share data with an application on their device to enable its functionality, while limiting the access of embedded trackers and third-party domains to that data.
\item \textit{Is it possible to create a profit sharing model between users and entities that sell users' data?}

The current data ecosystem involves various service providers, advertising agencies, and data gathering companies that profit significantly from sharing and selling individuals' personal information. Users are often overlooked in this system. Profit sharing models could be implemented to support the growth of these data-driven firms while also providing users with a portion of the profits.
\end{enumerate}

To enable users to take control of their data, it is necessary to establish a framework that grants them the authority to determine the collection and utilization of their data. This not only ensures a personalized experience with the service but also enables them to directly benefit from a portion of the revenue generated by the service provider.

\section{Conclusion}
This article conducts a thorough analysis of user profiling, delving into its security and privacy implications. It underscores the pros and cons of user profiling, stressing caution in granting access to personal information. We propose a taxonomy for building user profiles, offering a valuable resource for researchers. Identifying gaps in existing literature, we suggest future research directions. As a POC, we examine sensitive data leakage in applications across different categories. In regions without data protection laws, users face potential legal gaps in case of data breaches. It is crucial for users to weigh the benefits and risks of sharing information on platforms, especially in the absence of legal remedies. Users should stay vigilant, review terms and conditions, and make informed decisions about data sharing, given continuous corporate monitoring. Developing techniques that gather user profile information while respecting privacy is essential. Privacy-preserving methods for profiling can enhance user comfort with data collection. Ongoing efforts, like Apple's App Tracking Transparency in iOS 14.5, showcase steps toward safeguarding user privacy. A transparent data ecosystem that informs users about data sharing is crucial. Collaboration between commercial organizations and users is essential to build a digital world that leverages modern technologies while prioritizing privacy and security. Some countries, such as India, are introducing policies to protect their citizens.
\bibliography{sn-bibliography}

\end{document}